\pdfoutput=1
\documentclass[aps,prl,twocolumn,superscriptaddress,preprintnumbers,floatfix,nofootinbib]{revtex4-1}

\usepackage{graphicx}
\usepackage{amsmath}
\usepackage[caption=false]{subfig}
\usepackage{siunitx}
\usepackage{placeins}
\usepackage{color}
\usepackage{standalone}
\usepackage{dcolumn}
\usepackage{tensor}
\usepackage{bm}
\usepackage{microtype}
\usepackage{etoolbox}
\usepackage{amssymb}
\usepackage{mathrsfs}
\usepackage{accents}
\usepackage[normalem]{ulem}
\usepackage[dvipsnames]{xcolor}
\usepackage[colorlinks,urlcolor=NavyBlue,citecolor=NavyBlue,linkcolor=NavyBlue,pdfusetitle]{hyperref}
\usepackage[all]{hypcap}
\usepackage[inline]{enumitem}
\usepackage[utf8]{inputenc}

\newcommand{\beq}{\begin{equation}}
\newcommand{\eeq}{\end{equation}}

\newcommand*{\eq}[1]{Eq.~\eqref{eq:#1}}
\newcommand*{\fig}[1]{Fig.~\ref{fig:#1}}

\newcommand{\heading}[1]{\emph{#1}.}

\newcommand{\boldmu}{\boldsymbol{\mu}}
\newcommand{\boldn}{\mathbf{n}}
\newcommand{\boldd}{\mathbf{d}}
\newcommand{\bolds}{\mathbf{s}}
\newcommand{\boldSigma}{\boldsymbol{\Sigma}}

\newtoggle{commentsoff}
\togglefalse{commentsoff}

\ifdefined\nocomments
    \toggletrue{commentsoff}
\fi

\newtoggle{includeapp}
\toggletrue{includeapp}

\ifdefined\supplement
    \togglefalse{includeapp}
\fi

\iftoggle{commentsoff}{
  \newcommand*{\mi}[1]{}
  \newcommand*{\mg}[1]{}
  \newcommand*{\wf}[1]{}
  \newcommand*{\comment}[1]{}
  
  \newcommand*{\todo}[1]{}
  \newcommand*{\warn}[1]{}

  \newcommand*{\commentmark}[1]{}
}{
  \newcommand*{\mi}[1]{{\color{magenta} [{\bf MAX}: #1]}}
  \newcommand*{\mg}[1]{{\color{RedOrange} [{\bf MATT}: #1]}}
  \newcommand*{\wf}[1]{\textcolor{green}{\textbf{WILL}: #1}}
  
  \newcommand*{\commentmark}[1]{{\color{OliveGreen} [{\bf MARK}: #1]}}
  \newcommand*{\comment}[1]{{\color{blue} [{\bf NOTE}: #1]}}
  \newcommand*{\warn}[1]{{\color{red} [{\bf WARNING}: #1]}}
  \newcommand*{\todo}[1]{{\color{red} [{\bf TODO}: #1]}}

}

\graphicspath{{./fig/}}

\newcommand{\dcc}{LIGO-P1900135}

\newcommand{\mf}{M_f}
\newcommand{\chif}{\chi_f}
\newcommand{\tevent}{1126259462.423}
\newcommand{\tpeak}{t_{\rm peak}}

\begin{document}

\title{Testing the no-hair theorem with GW150914}

\author{Maximiliano Isi}
\email[]{maxisi@mit.edu}
\thanks{NHFP Einstein fellow}
\affiliation{
LIGO Laboratory, Massachusetts Institute of Technology, Cambridge, Massachusetts 02139, USA
}%

\author{Matthew~Giesler}
\affiliation{TAPIR, Walter Burke Institute for Theoretical Physics, California
Institute of Technology, Pasadena, CA 91125, USA}

\author{Will M. Farr}
\affiliation{Center for Computational Astrophysics, Flatiron Institute, 162 5th Ave, New York, NY 10010}
\affiliation{Department of Physics and Astronomy, Stony Brook University, Stony Brook NY 11794, USA}

\author{Mark~A.~Scheel}
\affiliation{TAPIR, Walter Burke Institute for Theoretical Physics, California
Institute of Technology, Pasadena, CA 91125, USA}

\author{Saul~A.~Teukolsky}
\affiliation{TAPIR, Walter Burke Institute for Theoretical Physics, California
Institute of Technology, Pasadena, CA 91125, USA}
\affiliation{Cornell Center for Astrophysics and Planetary Science, Cornell
University, Ithaca, New York 14853, USA}

\hypersetup{pdfauthor={Isi, Giesler, Farr, Scheel and Teukolsky}}

\date{\today}

\begin{abstract}
We analyze gravitational-wave data from the first LIGO detection of a binary black-hole merger (GW150914) in search of the ringdown of the remnant black hole.
Using observations beginning at the peak of the signal, we find evidence of the fundamental quasinormal mode and at least one overtone, both associated with the dominant angular mode ($\ell=m=2$), with $3.6\sigma$ confidence.
A ringdown model including overtones allows us to measure the final mass and spin magnitude of the remnant exclusively from postinspiral data, obtaining an estimate in agreement with the values inferred from the full signal.
The mass and spin values we measure from the ringdown agree with those obtained using solely the fundamental mode at a later time, but have smaller uncertainties.
Agreement between the postinspiral measurements of mass and spin and those using the full waveform supports the hypothesis that the GW150914 merger produced a Kerr black hole, as predicted by general relativity, and provides a test of the no-hair theorem at the ${\sim}10\%$ level.
An independent measurement of the frequency of the first overtone yields agreement with the no-hair hypothesis at the ${\sim 20}\%$ level.
As the detector sensitivity improves and the detected population of black hole mergers grows, we can expect that using
overtones will provide even stronger tests.
\end{abstract}

\maketitle

\heading{Introduction}
The coalescence of two astrophysical black holes consists of a long inspiral followed by a violent plunge, during which the full richness of spacetime dynamics comes into play.
The two objects merge, forming a single distorted black hole that rings down as it settles to a final stationary state.
Gravitational waves are emitted throughout the entire process, at each moment carrying information about the evolving source.
In general relativity, radiation from the ringdown stage takes the form of superposed damped sinusoids, corresponding to the quasinormal-mode oscillations of the final Kerr black hole~\cite{Vishveshwara1970b,Press1971,Teukolsky,ChandraDetweiler1975}.
The frequencies and decay rates of these damped sinusoids are uniquely determined by the final hole's mass $\mf$ and dimensionless spin magnitude $\chif$.
This is a consequence of the \emph{no-hair theorem}---the statement that mass and spin are the only two properties of astrophysical black holes in general relativity.%
\footnote{In general, black holes may also possess electric charge, but this is expected to be negligible for astrophysical objects.}
The ringdown spectrum is thus a fingerprint that identifies a Kerr black hole:
measuring the quasinormal modes from gravitational-wave observations would provide us with a unique laboratory to test general relativity and probe the true nature of remnants from compact-binary mergers, including testing the no-hair theorem~\cite{Echeverria:1989hg,Dreyer:2003bv,Berti:2005ys,Gossan:2011ha,Meidam:2014jpa,Berti:2015itd,Berti:2016lat,Baibhav:2017jhs,Baibhav:2018rfk}.
This program has been called \emph{black-hole spectroscopy}, in analogy to the spectroscopic study of atomic elements \cite{Dreyer:2003bv}.

Although LIGO \cite{aLIGO} and Virgo~\cite{Virgo} have already confidently detected gravitational waves from multiple binary-black-hole coalescences~\cite{gw150914,gw151226,o1bbh,gw170104,gw170608,gw170814,gwtc1:2018}, black hole spectroscopy has remained elusive \cite{gw150914_tgr,Nagar:2016iwa,Cabero:2017avf,Thrane:2017lqn,Brito:2018rfr,Carullo:2018sfu,Carullo:2019flw}.
This is because past analyses looked for the ringdown in data at late times after the signal peak, where the quasinormal modes are too weak to confidently characterize with current instruments.
The choice to focus on the late, weak-signal regime stemmed from concerns about nonlinearities surrounding the black hole merger, which were traditionally expected to contaminate the ringdown measurement at earlier times \cite{Gossan:2011ha, Kamaretsos:2011um, London:2014cma, Cabero:2017avf,Thrane:2017lqn, Carullo:2018sfu, Carullo:2019flw}.

Concerns about nonlinearities are, however, unfounded: the linear description can be extended to the full waveform following the peak of the gravitational wave strain \cite{Giesler:2019uxc}.
Rather than nonlinearities, times around the peak are dominated by ringdown \emph{overtones}---the quasinormal modes with the fastest decay rates, but also the highest amplitudes near the waveform peak \cite{Giesler:2019uxc,Buonanno:2006ui}.
Indications of this can be found in the waveform modeling literature, with overtones an integral part of earlier equivalent one-body models~\cite{Pan:2013rra,Taracchini:2013rva,Babak:2016tgq} (although later abandoned, c.f.~\cite{Bohe:2016gbl}).
Yet, with a few exceptions~\cite{Baibhav:2017jhs,Brito:2018rfr}, previous ringdown analyses have neglected overtones, under the assumption that their contribution to the signal should always be marginal \cite{Gossan:2011ha, gw150914_tgr,Bhagwat:2016ntk,Nagar:2016iwa,Cabero:2017avf,Thrane:2017lqn,Carullo:2018sfu,Carullo:2019flw}.
As a consequence, these studies ignored important signal content and were unable to extract multiple ringdown modes.

\begin{figure}[btp]
\includegraphics[width=\columnwidth,clip=true]{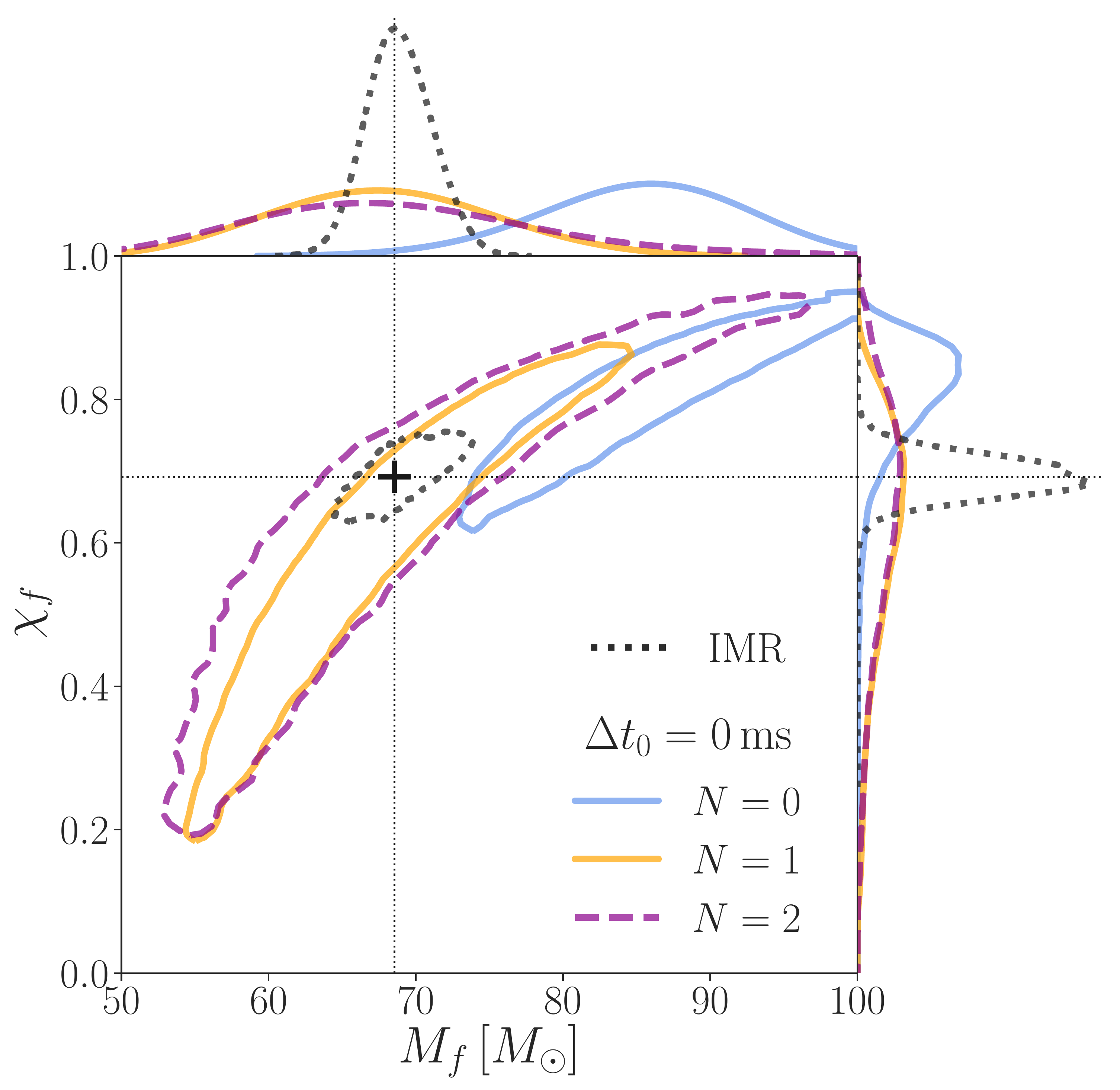}
\caption{
    Remnant parameters inferred with different number of overtones, using data starting at peak strain amplitude.
    Contours represent 90\%-credible regions on the remnant mass ($M_f$) and dimensionless spin magnitude ($\chi_f$), obtained from the Bayesian analysis of GW150914.
    The inference model is that of \eq{qnm}, with different number of overtones $N$: 0 (solid blue), 1 (solid yellow), 2 (dashed purple). In all cases, the analysis uses data starting at peak strain ($\Delta t_0 = t_0 - t_{\rm peak} = 0$).
    Amplitudes and phases are marginalized over.
    The black contour is the 90\%-credible region obtained from the full IMR waveform, as described in the text.
    The intersection of the dotted lines marks the peak of this distribution ($M_f = 68.5 M_\odot$, $\chi_f=0.69$).
    The top and right panels show 1D posteriors for $M_f$ and $\chi_f$ respectively.
    The linear quasinormal mode models with $N > 0$ provide measurements of the mass and spin consistent with the full IMR waveform, in agreement with general relativity.
}
\label{fig:contours_n}
\end{figure}

\begin{figure}[btp]
\includegraphics[width=\columnwidth,clip=true]{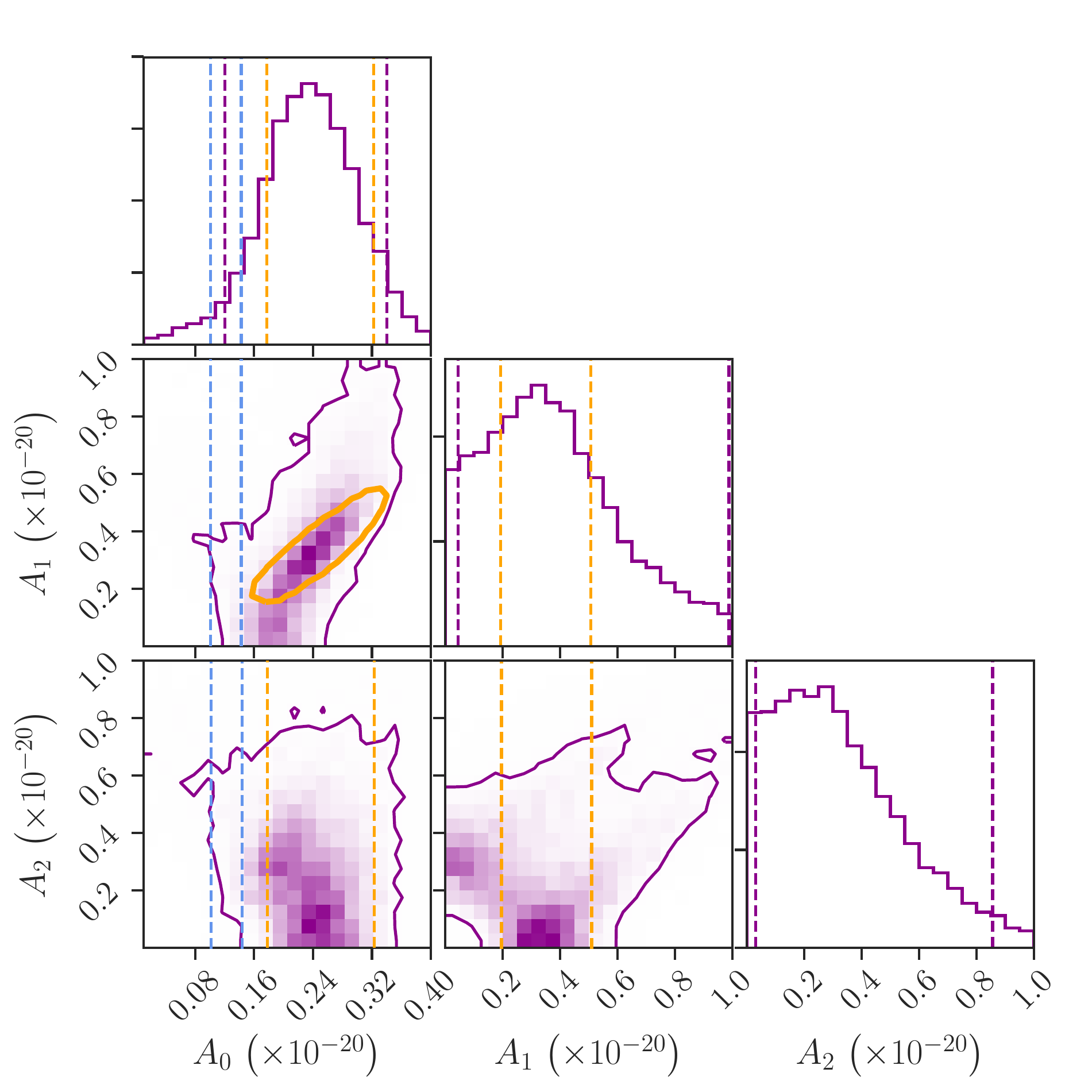}
\caption{
Measured quasinormal-mode amplitudes for a model with the fundamental mode and two overtones ($N=2$).
The purple colormap represents the joint posterior distribution for the three amplitudes in the $N=2$ model: $A_0$, $A_1$, $A_2$, as defined in \eq{qnm}.
The solid curves enclose 90\% of the probability mass.
A yellow curve in the $A_0$--$A_1$ plane, as well as corresponding yellow dashed lines, represents the 90\%-credible measurement of the amplitudes assuming $N=1$.
Similarly, blue dashed lines give the 90\%-credible measurement of $A_0$ assuming $N=0$.
All amplitudes are defined at $t=t_{\rm peak}$, where all fits here are carried out ($\Delta t_0=0$).
Values have been rescaled by a constant to correspond to the strain measured by the LIGO Hanford detector.
Assuming $N=1$, the mean of the $A_1$ marginalized posterior lies {3.6} standard deviations away from zero, i.e.~$A_1=0$ is disfavored at {$3.6\sigma$}.
Assuming $N=2$, $A_1=A_2=0$ is disfavored with 90\% credibility.
}
\label{fig:allamps}
\end{figure}

The inclusion of overtones enables us to perform a multimodal spectroscopic analysis of a black-hole
ringdown, which we apply to LIGO data from the GW150914 event~\cite{gw150914} (\fig{contours_n}).
We rely on overtones of the $\ell=m=2$ angular mode to measure the remnant mass and spin from data starting
at the peak of the signal, assuming first that quasinormal modes are as
predicted for a Kerr black hole within general relativity. We find the least-damped
(`fundamental') mode and at least one overtone with {$3.6\sigma$ confidence} (\fig{allamps}).
At least one overtone, in addition to the fundamental, is needed to describe the waveform near the peak amplitude.
This agrees with our expectations from \cite{Giesler:2019uxc} given the signal-to-noise ratio of GW150914.

\begin{figure}[btp]
\includegraphics[width=\columnwidth,clip=true]{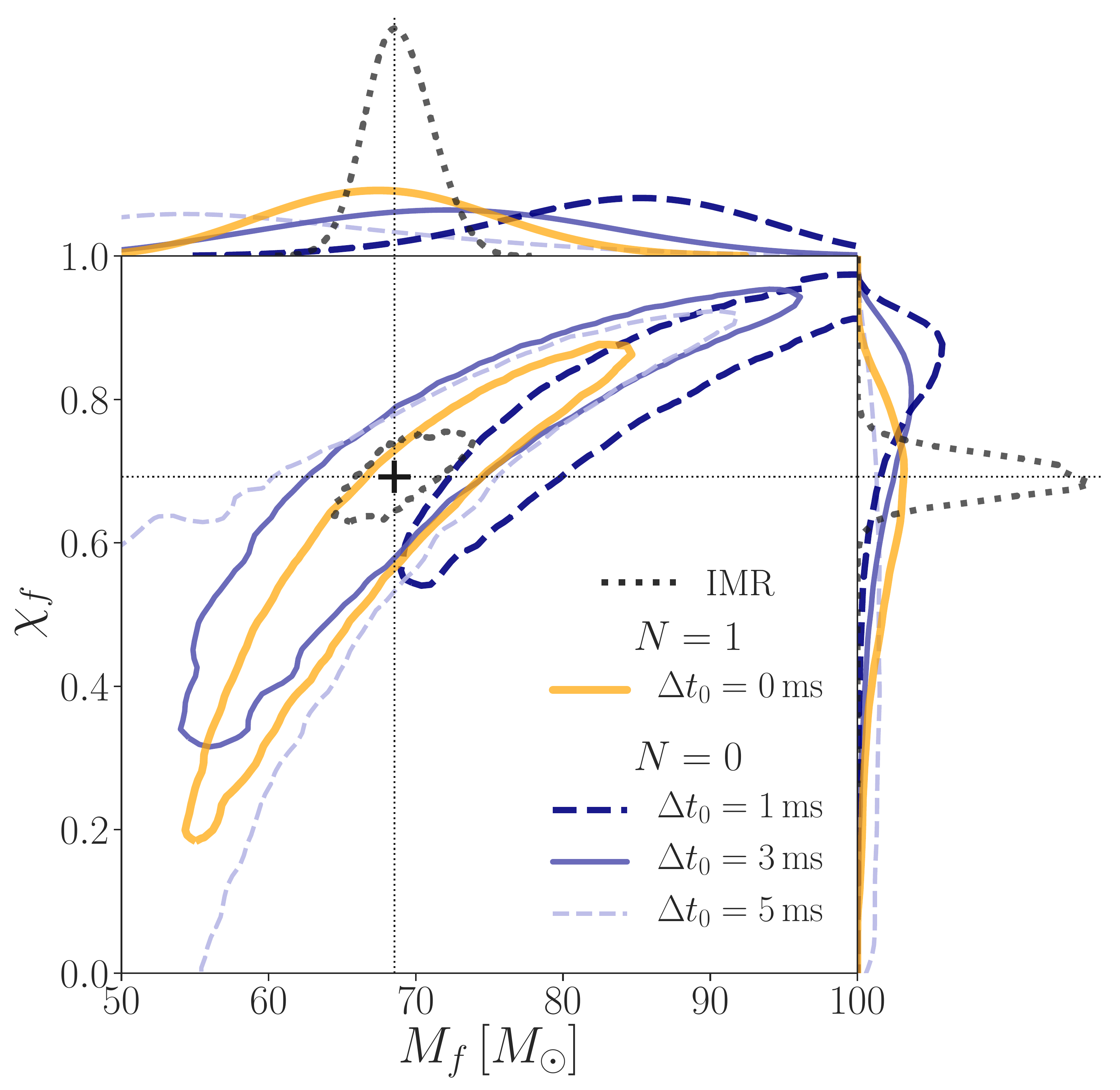}
\caption{
    Remnant parameters inferred only from the fundamental mode, using data
    starting at different times after the peak.
    Contours represent 90\%-credible regions on the remnant mass
    ($M_f$) and dimensionless spin magnitude ($\chi_f$), obtained from the Bayesian
    analysis of GW150914. For the blue contours, the inference model included
    no overtones ($N=0$) and used data starting at different times after the
    peak: $\Delta t_0 = t_0-t_{\rm peak} \in [1,\,3,\,5]\,\mathrm{ms}$
    . For the yellow contour, the analysis was conducted with
    one overtone ($N=1$) starting at the peak ($\Delta t_0=0$), as in
    Fig.~\ref{fig:contours_n}.
    Amplitudes and phases are marginalized over.
    The black contour is the 90\%-credible region obtained from the full IMR waveform, as described in the text.
    The intersection of the dotted lines marks the peak of this distribution ($M_f = 68.5 M_\odot$, $\chi_f=0.69$).
    The top and right panels show 1D posteriors for $M_f$ and $\chi_f$ respectively.
    Around $\Delta t_0 = 3\, \mathrm{ms}$, the overtones have become unmeasurable and only the fundamental mode remains; consequently, at that time $N=0$ returns a measurement of the final mass and spin consistent with both the full IMR waveform and the $N>0$ models at the peak, in agreement with general relativity.
}
\label{fig:contours_t0}
\end{figure}

Assuming the remnant is a Kerr black hole, frequencies and damping rates of the fundamental mode and one overtone imply a detector-frame mass of {$\left(68\pm7\right) M_\odot$} and a dimensionless spin magnitude of {$0.63\pm0.16$}, with {68}\% credibility.
This is the best constraint on the remnant mass and spin obtained in this work.
This measurement agrees with the one obtained from the fundamental mode alone beginning $3 \, \mathrm{ms}$ after the waveform peak amplitude (Figures \ref{fig:contours_n} and \ref{fig:contours_t0}) \citep{TheLIGOScientific:2016src}.
It also agrees with the mass and spin inferred from the full waveform using fits to numerical relativity.
The fractional difference between the best-measured combination of mass and spin%
\footnote{That is, the measurement of the linear combination of $\mf$ and $\chif$ corresponding to the principal component of the posterior distribution with the smallest associated eigenvalue.}
at the peak with one overtone and the same combination solely with the fundamental $3 \, \mathrm{ms}$ after the peak is {$(0\pm10)\%$}.
This is evidence at the {${\sim}10\%$ level} that GW150914 did result in a Kerr black hole as
predicted by general relativity, and that the postmerger signal is in agreement with the no-hair theorem.
Similarly, the fractional difference between the best-measured combination of mass and spin at the peak with one overtone and the same
combination using the full waveform is {$\left(7 \pm 7\right) \%$}.

\begin{figure}[btp]
\includegraphics[width=\columnwidth,clip=true]{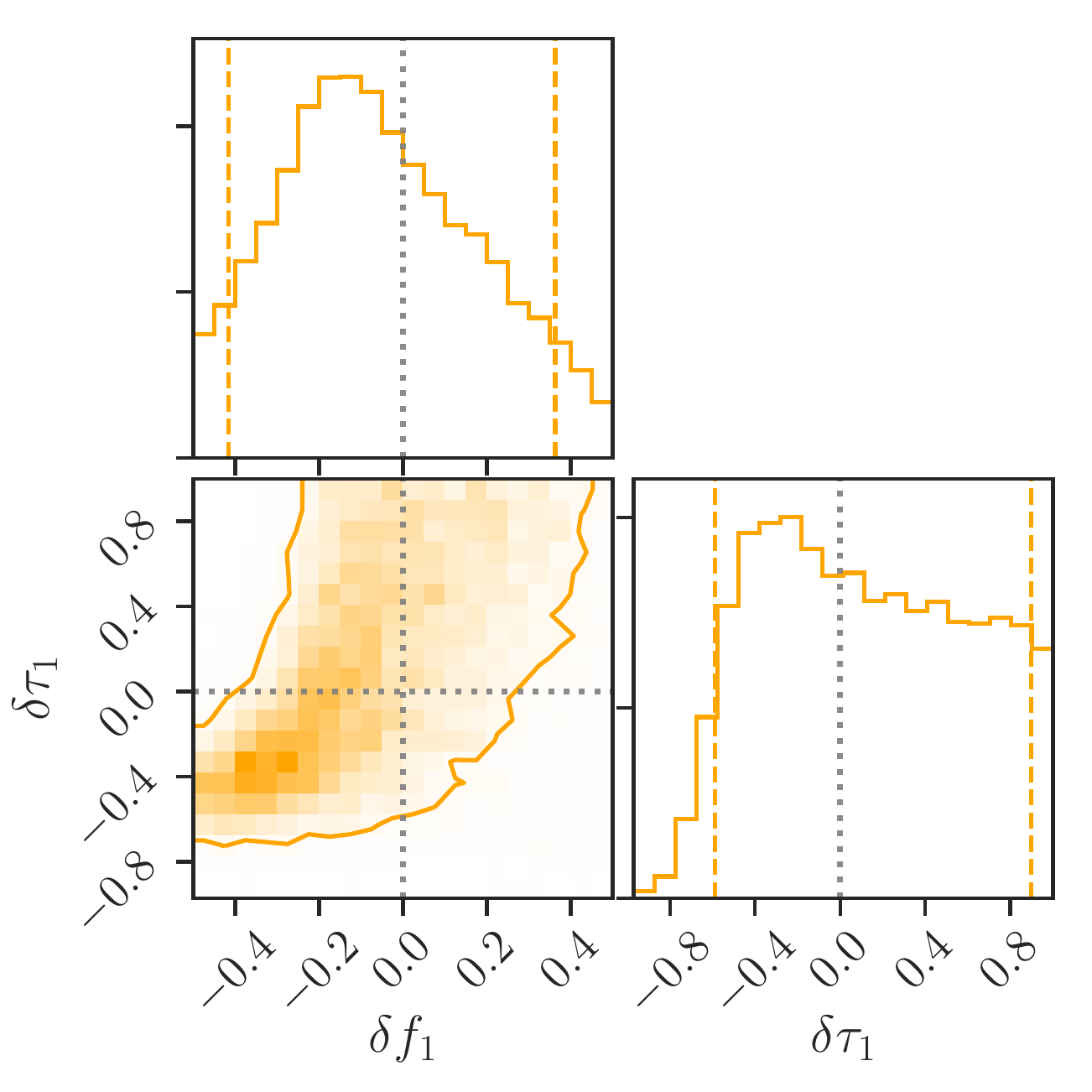}
\caption{
Measurement of the frequency and damping time of the first overtone, using data starting at the peak.
The colormap represents the posterior distribution of the fractional deviations $\delta f_1$ and $\delta\tau_1$ away from the no-hair value $\delta f_1=\delta\tau_1=0$ (gray dotted lines).
The solid contour and dashed vertical lines enclose 90\% of the posterior probability.
All other parameters, including $M_f$ and $\chi_f$ have been marginalized away.
Fixing $\delta f_1=\delta\tau_1 = 0$ recovers the $N=1$ analysis in Figs.~\ref{fig:contours_n} and \ref{fig:contours_t0}.
}
\label{fig:deltaf1}
\end{figure}

Traditional proposals for black-hole spectroscopy require frequency measurements for two or more quasinormal modes \cite{Dreyer:2003bv}.
In that spirit, we also consider a single-overtone model that allows the overtone frequency and damping time to deviate from the Kerr prediction for any given mass and spin.
This enables us to evaluate agreement of the observed ringdown spectrum with the prediction for a perturbed Kerr black hole, regardless of the specific properties of the remnant.
From analysis of data starting at peak strain, we find the spectrum to be in agreement with the no-hair hypothesis to within ${\sim}20\%$, with $68\%$ credibility (\fig{deltaf1}).
This is a test of the no-hair theorem based purely on the postinspiral regime.

\heading{Method}
Each quasinormal mode has a frequency $\omega_{\ell m n}$ and a damping time $\tau_{\ell m n}$, where $n$ is the `overtone' index and $(\ell,m)$ are indices of spin-weighted angular harmonics that describe the angular dependence of the mode.
We focus on the fundamental and overtones of the dominant $\ell=m=2$ spin-weighted spherical harmonic of the strain.%
\footnote{The spin-weighted \emph{spheroidal}
harmonics form the natural basis that arises in perturbation theory~\cite{Teukolsky:1972my,Teukolsky,Press:1973}.
These functions are equivalent to the spin-weighted spherical harmonics in the limit of zero spin. For $\chi_f>0$, the spin-weighted spheroidal harmonics can be written as superpositions of the spin-weighted spherical harmonics of the same $m$, but different $\ell$~\cite{Press:1973,Berti:2014fga}.
The effect of this mixing on the dominant $\ell=m=2$ spin-weighted spherical mode is negligible for a GW150914-like system~\cite{Giesler:2019uxc}.}
This is the only angular harmonic expected to be relevant for GW150914 \cite{Abbott:2016apu,Abbott:2016wiq}.%
\footnote{Dedicated studies have found no evidence of higher angular harmonics in the late ringdown of GW150914 \cite{Carullo:2019flw}.}
For ease of notation, we generally drop the $\ell$ and $m$ indices, retaining only the overtone index $n$.
The $\ell=m=2$ mode of the parametrized ringdown strain ($h=h_+ - i h_\times$) can be written as a sum of damped sinusoids~\cite{Vishveshwara1970b,Press1971,Teukolsky,ChandraDetweiler1975},
\beq\label{eq:qnm}
  h_{22}^{N}(t) = \sum\limits_{n=0}^{N} A_n \exp\left[-i \left(\omega_n t + \phi_n\right) - t/\tau_n\right] ,
\eeq
for times $t$ greater than some start time $t_0$, where $\Delta t = t-t_0$.
The overtone index $n$ orders the different modes by decreasing damping time $\tau_n$, so that $n=0$ denotes the longest-lived mode.
$N$ is the index of the highest overtone included in the model, which in this work will be $N\leq 2$.
Importantly, higher $n$ does not imply a higher frequency $\omega_n$; rather, the opposite is generally true.
All frequencies and damping times are implicit functions of the remnant mass and spin magnitude ($\mf,\, \chif$), and can be computed from perturbation theory \cite{Leaver1985,Berti2009,BertiWebsite}.
The amplitudes $A_n$ and phases $\phi_n$ encode the degree to which each overtone is excited as the remnant is formed and cannot be computed within perturbation theory, so we treat them as free parameters in our fit.

We use the model in \eq{qnm} to carry out a Bayesian analysis of LIGO Hanford and LIGO Livingston data for GW150914~\cite{gw150914,gwtc1:2018,gwosc}.
For any given start time $t_0$, we produce a posterior probability density over the space of remnant mass and spin magnitude, as well as the amplitudes and phases of the included overtones.
We parametrize start times via $\Delta t_0 = t_0 - \tpeak$, where $\tpeak = \tevent$ GPS refers to the inferred signal peak at the LIGO Hanford detector \cite{gw150914_pe,gw150914_tgr}.
We define the likelihood in the time domain in order to explicitly exclude all data before $t_0$.
We place uniform priors on $\left(\mf,\, \chif,\, A_n,\, \phi_n\right)$, with a restriction to corotating modes ($\omega_n>0$).
All overtones we consider share the same $\ell=m=2$ angular dependence, allowing us to simplify the handling of antenna patterns and other subtleties.
Details specific to our implementation are provided in the supplementary material.

We compare our ringdown-only measurements of the remnant mass and spin magnitude to those obtained from the analysis of the full inspiral-merger-ringdown (IMR) signal.
To do so, we rely on fitting formulas based on numerical relativity to translate measured values of the binary mass ratio $q$ and component spins $(\vec{\chi}_1, \vec{\chi}_2)$ into expected remnant parameters~\cite{Varma:2018aht,Blackman:2017pcm}.
We use posterior samples on the binary parameters made available by the LIGO and Virgo collaborations \cite{gwtc1:2018,GWOSC:GWTC}, marginalizing over unavailable component-spin angles.

We consider explicit deviations from the Kerr spectrum by allowing the frequency and damping time of the first overtone to differ from the no-hair values.
Under this modified $N=1$ model, the overtone angular frequency becomes $\omega_1=2\pi f^{\rm (GR)}_1\left(1+\delta f_1\right)$, with $\delta f_1$ a fractional deviation away from the Kerr frequency $f^{\rm (GR)}_1$ for any given $\mf$ and $\chif$.
Similarly, the damping time is allowed to vary by letting $\tau_1 = \tau^{\rm (GR)}_1\left(1+\delta\tau_1\right)$.
Fixing $\delta f_1=\delta\tau_1=0$ recovers the regular $N=1$ analysis.
We may then compute the relative likelihood of the no-hair hypothesis by means of the Savage-Dickey density ratio \cite{Verdinelli1995}.

\heading{Results}
\fig{contours_n} shows the 90\%-credible regions for the remnant mass (abscissa) and spin magnitude (ordinate) obtained by analyzing data starting at $\tpeak$ with different numbers of overtones ($N=0,1,2$) in the ringdown template of \eq{qnm}.
The quasinormal-mode amplitudes and phases have been marginalized over.
For comparison, we also show the 90\%-credible region inferred from the full IMR signal, as explained above.
If the remnant is sufficiently well described as a perturbed Kerr black hole, and if general relativity is correct, we expect the ringdown and IMR measurements to agree.
As expected, this is not the case if we assume the ringdown is composed solely of the longest-lived mode ($N=0$), in which case we obtain a biased estimate of the remnant properties.
In contrast, the ringdown and IMR measurements begin to agree with the addition of one overtone ($N=1$).
This is expected from previous work suggesting that, given the network signal-to-noise ratio of GW150914 (${\sim}14$ in the post-peak region, for frequencies ${>}154.7\, \mathrm{Hz}$), we should be able to resolve only one mode besides the fundamental \cite{Giesler:2019uxc}.

Indeed, a ringdown model with two overtones ($N=2$) does not lead to further improvement in the mass and spin measurement.
On the contrary, the 90\%-credible region obtained with $N=2$ is slightly broader than the one with $N=1$, as might be expected from the two additional free parameters ($A_2,\, \phi_2$).
This is because the analysis is unable to unequivocally identify the second overtone in the data, as shown by the amplitude posteriors in \fig{allamps}.
The $N=2$ posterior supports a range of values for $A_1$ and $A_2$, but excludes $A_1=A_2=0$ with 90\% credibility (center panel in bottom row of \fig{allamps}).
The joint posterior distribution on $A_1$ and $A_2$ tends to favor the first overtone at the expense of the second: the maximum a posteriori waveform scarcely includes any contribution from $n=2$, and favors a value of $A_1$ in agreement with the $N=1$ posterior (yellow traces in \fig{allamps}).

We next compare measurements carried out with overtones at the peak with measurements without overtones after the peak.
\fig{contours_t0} shows 90\%-credible regions for the remnant mass and spin magnitude obtained with the fundamental mode ($N=0$) at different times after $\tpeak$ ({$\Delta t_0 \in [1,\,3,\,5]\,\mathrm{ms}$}).
As the overtones die out, the fundamental mode becomes a better model for the signal.
We find that the $N=0$ contour coincides with the IMR measurement ${\sim}3\,$ms after the peak, in agreement with \cite{gw150914_tgr}.
However, the uncertainty in this measurement is larger than for the $N=1$ contour at the peak (also shown for reference).
This can be attributed to the exponential decrease in signal-to-noise ratio for times after the peak.

Finally, we allow the first-overtone frequency and damping time to float around the no-hair values in an $N=1$ model.
As in \fig{contours_n}, we analyze data starting at the inferred peak of the strain.
\fig{deltaf1} shows the resulting marginalized posterior over the fractional frequency and damping time deviations ($\delta f_1$ and $\delta\tau_1$ respectively).
With 68\% credibility, we measure $\delta f_1 = -0.05 \pm 0.2$.
To that level of credibility, this establishes agreement with the no-hair hypothesis ($\delta f_1 = 0$) at the 20\% level.
The damping time is largely unconstrained in the $-0.06 \lesssim \delta\tau_1 \lesssim 1$ range.
This has little impact on the frequency measurement, which is unaffected by setting $\delta\tau_1=0$.
We find that the ratio of marginal likelihoods (the Bayes factor) between the no-hair model ($\delta f_1 = \delta \tau_1 = 0$) and our floating frequency and damping time model is 1.75.

\heading{Discussion and prospects}
A linearly perturbed Kerr black hole radiates gravitational waves in the form of damped sinusoids, with specific frequencies and decay rates determined exclusively by the hole's mass and spin.
For any given angular harmonic, the quasinormal modes can be ordered by decreasing damping time through an overtone index $n$, with $n=0$ denoting the longest-lived mode (also known as the `fundamental').
Although modes of all $n$ contribute to the linear description, the fundamental has long been the only one taken into account in observational studies of the ringdown, with overtones virtually ignored \cite{gw150914_tgr,Nagar:2016iwa,Cabero:2017avf,Thrane:2017lqn,Carullo:2018sfu,Carullo:2019flw}.
Yet, these short-lived modes can dominate the gravitational wave signal for times around the peak and are an essential part of the ringdown \cite{Giesler:2019uxc,Buonanno:2006ui}.
We demonstrate this with a multimode analysis of the GW150914 ringdown.

Making use of overtones, we extract information about the GW150914 remnant using only postinspiral data, starting at the peak of the signal (\fig{contours_n}).
We find evidence of the fundamental mode plus at least one overtone (\fig{allamps}), and obtain a 90\%-credible measurement of the remnant mass and spin magnitude in agreement with that inferred from the full waveform.
This measurement is also consistent with the one obtained using solely the fundamental mode at a later time, but has reduced uncertainties (\fig{contours_t0}).

The agreement between all measurements is evidence that, beginning as early as the signal peak, a far-away observer cannot distinguish the source from a linearly perturbed Kerr background
with a fixed mass and spin, i.e., we do not observe nonlinearities in this regime.
The agreement between the IMR and postmerger estimates implies that the data agree with the full prediction of general relativity.
This is similar to the consistency test between inspiral and merger-ringdown \cite{Ghosh:2016qgn,Ghosh:2017gfp}, but relies on a manifestly linear description of the postinspiral signal.
More specifically, it validates the prediction for the final state of a collision between two black holes.

With the identification of multiple ringdown modes, this is also a step toward the goal of black hole spectroscopy.
The agreement between postinspiral measurements with two different sets of modes (\fig{contours_t0}) supports the hypothesis that GW150914 produced a Kerr black hole as described by general relativity.
Moreover, we constrain deviations away from the no-hair spectrum by allowing the overtone frequency and damping time to vary freely (\fig{deltaf1}).
This is equivalent to independently measuring the frequencies of the fundamental and first overtone, and establishing their consistency with the Kerr hypothesis.

Future studies of black-hole ringdowns relying on overtones could potentially allow us to identify black-hole mimickers and probe the applicability of the no-hair theorem with high precision, even with existing detectors.
Such advances will be facilitated by improvements in our understanding of how the overtones are sourced, so that we can predict the amplitudes and phases from the binary properties.
This would reduce the dimensionality of the problem and lead to more specific predictions from general relativity.

\begin{acknowledgments}
\heading{Acknowledgments}
We thank Aaron Zimmerman for valuable feedback.
We thank Gregorio Carullo, Walter del Pozzo, and John Veitch for discussions of
their paper on and methods for time-domain analysis \citep{Carullo:2019flw}.
We thank Alessandra Buonanno for clarifications on past use of quasinormal and pseudo-quasinormal ringdown modes in waveform modeling.
M.I.\ is supported by NASA through the NASA Hubble Fellowship
grant No.\ HST-HF2-51410.001-A awarded by the Space Telescope
Science Institute, which is operated by the Association of Universities
for Research in Astronomy, Inc., for NASA, under contract NAS5-26555.
M.I.\ is a member of the LIGO Laboratory.
LIGO was constructed by the California Institute of Technology and
Massachusetts Institute of Technology with funding from the National
Science Foundation and operates under cooperative agreement PHY-0757058.
M.G.\ and M.S.\ are supported by the Sherman Fairchild Foundation and NSF
grants PHY-1708212 and PHY-1708213 at Caltech.
S.T.\ is supported in part by the Sherman Fairchild Foundation and by NSF Grants
PHY-1606654 and ACI-1713678 at Cornell.
The Flatiron Institute is supported by the Simons Foundation.
This research has made use of data, software and/or web tools obtained from the Gravitational Wave Open Science Center \cite{Vallisneri:2014vxa,gwosc}, a service of the LIGO Laboratory, the LIGO Scientific Collaboration and the Virgo Collaboration.
This paper carries LIGO document number \dcc{}.
\end{acknowledgments}

\bibliography{nr,ligo}

\appendix

\iftoggle{includeapp}{
    \newcommand{\includeapp}{}
    \section{APPENDIX: TECHNICAL DETAILS}
    \ifdefined\includeapp
    
\else
    \input{preamble}
    \togglefalse{includeapp}

    \begin{document}

    \title{Supplemental material for\\``Testing the no-hair theorem with GW150914''}

\author{Maximiliano Isi}
\email[]{maxisi@mit.edu}
\thanks{NHFP Einstein fellow}
\affiliation{
LIGO Laboratory, Massachusetts Institute of Technology, Cambridge, Massachusetts 02139, USA
}%

\author{Matthew~Giesler}
\affiliation{TAPIR, Walter Burke Institute for Theoretical Physics, California
Institute of Technology, Pasadena, CA 91125, USA}

\author{Will M. Farr}
\affiliation{Center for Computational Astrophysics, Flatiron Institute, 162 5th Ave, New York, NY 10010}
\affiliation{Department of Physics and Astronomy, Stony Brook University, Stony Brook NY 11794, USA}

\author{Mark~A.~Scheel}
\affiliation{TAPIR, Walter Burke Institute for Theoretical Physics, California
Institute of Technology, Pasadena, CA 91125, USA}

\author{Saul~A.~Teukolsky}
\affiliation{TAPIR, Walter Burke Institute for Theoretical Physics, California
Institute of Technology, Pasadena, CA 91125, USA}
\affiliation{Cornell Center for Astrophysics and Planetary Science, Cornell
University, Ithaca, New York 14853, USA}

\hypersetup{pdfauthor={Isi, Giesler, Farr, Scheel and Teukolsky}}

    \date{\today}
    \maketitle

    \section{TECHNICAL DETAILS}
\fi

\heading{Time-domain likelihood}
The signal model used in this analysis has a sharp transition in the
time domain.  We do not wish to incorporate any data from the detectors before
our model begins so that we avoid bias; equivalently, we want to assume that we
are infinitely uncertain about the gravitational wave signal before the start
point of our signal model.  This requires special treatment compared to the
standard LIGO data analysis
\cite{TheLIGOScientific:2016wfe,TheLIGOScientific:2016src}. Other approaches to
quasi-normal mode extraction treat the data similarly \cite{Carullo:2019flw}
except for the assumption of periodicity, enforced by tapering, which is absent
in our treatment (see below).

As in previous analyses of GW150914, we assume that the detector measures a
discrete data stream $\boldd$ that contains our signal $\bolds$ contaminated by additive, Gaussian
noise $\boldn$.  At time $t_i$,
\begin{equation}
  d_i = h( t_i ) + n_i\, ,
\end{equation}
with the noise time series $\boldn{}$ having a multivariate normal distribution,
\begin{equation}
  \label{eq:noise-multivariate-gaussian}
  \boldn{} \sim \mathcal{N} (\boldmu{}, \boldSigma{} )\, ,
\end{equation}
with mean $\boldmu{}$ and covariance matrix $\boldSigma{}$.  The actual noise in
the detector has very large low-frequency components, while our signal lives
primarily at high ($\gtrsim 100 \, \mathrm{Hz}$) frequencies \cite{Berti2009,BertiWebsite}.
To reduce the low-frequency components of the noise,
 we first apply a fourth-order, high-pass Butterworth filter with a roll-on frequency of 20 Hz.
 After the filter, the data are very close to zero mean, so we assume $\boldmu{} = \mathbf{0}$.

The distribution of the noise implies that the log-likelihood function (i.e., the
distribution of data $\mathbf{d}$ conditioned on a signal $\mathbf{h}$) is
\begin{align}
  \label{eq:mg-log-likelihood}
  \log p\left( \mathbf{d} \mid \mathbf{h} \right) = &-\frac{1}{2} \left( \mathbf{d} - \mathbf{h} \right)^T \boldSigma{}^{-1} \left( \mathbf{d} - \mathbf{h} \right) \nonumber \\
  &- \frac{1}{2} \log \det \boldSigma - \frac{K}{2} \log 2 \pi\, ,
\end{align}
where $K$ is the total number of samples.

We assume that the noise in the detector is stationary, so that the covariance
matrix takes a special (Toeplitz) form where the $ij$ component depends only on
the time separation between samples $i$ and $j$:
\begin{equation}
  \Sigma_{ij} = \left\langle n_i n_j \right\rangle = \rho\left( \left| i - j \right| \right),
\end{equation}
where $\rho$ is the autocovariance function,
\begin{equation}
  \rho\left( k \right) = \left \langle n_i n_{i+k} \right\rangle.
\end{equation}
(The expectation above runs over all times, $i$.)  The assumptions of stationary
Gaussian noise have been checked for GW150914 specifically
\cite{TheLIGOScientific:2016zmo,TheLIGOScientific:2016uux,TheLIGOScientific:2016src,TheLIGOScientific:2016wfe}
and the LIGO events in general \cite{TheLIGOScientific:2016pea,gwtc1:2018}.

The standard LIGO data analysis imposes an additional assumption that
\begin{equation}
  \rho\left( k \right) = \rho\left( K - k \right)
\end{equation}
for $0 \leq k < K$.  This
``circularity'' assumption is appropriate for data that are periodic with
period $K$; periodicity is typically enforced by tapering the data segment at
the beginning and end \cite{Veitch:2015,TheLIGOScientific:2016wfe}.  The
benefit of this assumption is that a circular Toeplitz matrix is diagonal in the
Fourier basis (i.e.\ stationary periodic noise has statistically-independent
Fourier components), and therefore the matrix-inversion step in the
log-likelihood reduces to a sum over independent frequency components.  Such a
likelihood can also be computed directly in the time domain
\cite{Carullo:2019flw}.

A taper is not appropriate for our data analysis since we wish to ignore data
from times before the peak of the waveform, where our signal model begins. We do
not have data before the peak in which to implement a taper; and tapering past
the peak would significantly reduce our signal.  Happily, fast and stable
algorithms exist for solving linear equations with a Toeplitz structure
\cite{Levinson:1947,Durbin:1960,Jones:2001scipy}, so a direct implementation of
our likelihood in Eq.\ \eqref{eq:mg-log-likelihood} is not too costly.

We estimate the autocovariance function by the empirical autocovariance of
$64\, \mathrm{s}$ of off-source data, after high-pass filtering as above.
This is analogous to the Welch method for estimating power spectral densities
in the frequency domain \cite{Welch:1967use} used by the standard LIGO analyses
\cite{Veitch:2015,TheLIGOScientific:2016wfe}.  Our analysis is based on data at
a sample rate of $2048 \, \mathrm{Hz}$, beginning at the peak signal amplitude
at \tevent{} GPS and running for $0.5 \, \mathrm{s}$.
The autocovariance estimate is truncated to that same duration.

\heading{Other details}
\ifdefined\includeapp
We handle polarizations by projecting the complex-valued strain in \eq{qnm} onto each LIGO detector by means of the corresponding antenna patterns.
\else
We handle polarizations by projecting the complex-valued strain in Eq.~(1) of the main text onto each LIGO detector by means of the corresponding antenna patterns.
\fi
To do so, we assume the source of GW150914 had right ascension $\alpha = 1.95 $ rad and declination $\delta = -1.27$ rad, with polarization angle $\psi = 0.82$ rad and inclination $\iota = \pi$ rad.
These parameters are consistent with the maximum a posteriori estimates inferred for GW150914 \cite{gw150914,gw150914_pe,GWOSC:GWTC}.
We also time-shift the LIGO Livingston data by the corresponding arrival-time delay of 7 ms \cite{gw150914,gw150914_pe}, so as to align the signal at the two detectors.
As noted in the main text, we may make these simplifications because all rindgown modes we consider are subject to the same angular dependence ($\ell=m=2$).
A version of this analysis with the more simplified approach of \cite{gw150914_tgr,bayesian_qnm} yields compatible results.

Our priors are such that quasinormal-mode amplitudes $A_n$ are allowed to vary in the range $[0,\,2.5\times10^{-19}]$, an arbitrary range found to offer full support to the posterior in all cases.
The corresponding phases $\phi_n$ are unrestricted in the full range $[0,\,2\pi]$.
For computational efficiency, we internally parameterize the amplitude and phase of each mode using the two quadratures $c_n = A_n \cos \phi_n$ and $s_n = A_n \sin \phi_n$, but set priors uniform in $A_n$ and $\phi_n$.
The remnant mass $\mf$ is allowed to vary within $[50,\,100]\,M_\odot$, while the dimensionless spin magnitude $\chif$ varies within $[0,\,1]$. 
When considering explicit deviations from the no-hair theorem, we set uniform priors such that $-0.5 < \delta f_1 < 0.5$ and $-1 < \delta \tau_1 < 1$.
In all cases, samples are drawn from the posterior using \texttt{kombine} \cite{kombine}.

\ifdefined\includeapp
\else
    \bibliography{nr,ligo}

\begin{thebibliography}{67}%
\makeatletter
\providecommand \@ifxundefined [1]{%
 \@ifx{#1\undefined}
}%
\providecommand \@ifnum [1]{%
 \ifnum #1\expandafter \@firstoftwo
 \else \expandafter \@secondoftwo
 \fi
}%
\providecommand \@ifx [1]{%
 \ifx #1\expandafter \@firstoftwo
 \else \expandafter \@secondoftwo
 \fi
}%
\providecommand \natexlab [1]{#1}%
\providecommand \enquote  [1]{``#1''}%
\providecommand \bibnamefont  [1]{#1}%
\providecommand \bibfnamefont [1]{#1}%
\providecommand \citenamefont [1]{#1}%
\providecommand \href@noop [0]{\@secondoftwo}%
\providecommand \href [0]{\begingroup \@sanitize@url \@href}%
\providecommand \@href[1]{\@@startlink{#1}\@@href}%
\providecommand \@@href[1]{\endgroup#1\@@endlink}%
\providecommand \@sanitize@url [0]{\catcode `\\12\catcode `\$12\catcode
  `\&12\catcode `\#12\catcode `\^12\catcode `\_12\catcode `\%12\relax}%
\providecommand \@@startlink[1]{}%
\providecommand \@@endlink[0]{}%
\providecommand \url  [0]{\begingroup\@sanitize@url \@url }%
\providecommand \@url [1]{\endgroup\@href {#1}{\urlprefix }}%
\providecommand \urlprefix  [0]{URL }%
\providecommand \Eprint [0]{\href }%
\providecommand \doibase [0]{http://dx.doi.org/}%
\providecommand \selectlanguage [0]{\@gobble}%
\providecommand \bibinfo  [0]{\@secondoftwo}%
\providecommand \bibfield  [0]{\@secondoftwo}%
\providecommand \translation [1]{[#1]}%
\providecommand \BibitemOpen [0]{}%
\providecommand \bibitemStop [0]{}%
\providecommand \bibitemNoStop [0]{.\EOS\space}%
\providecommand \EOS [0]{\spacefactor3000\relax}%
\providecommand \BibitemShut  [1]{\csname bibitem#1\endcsname}%
\let\auto@bib@innerbib\@empty
\bibitem [{\citenamefont {Vishveshwara}(1970)}]{Vishveshwara1970b}%
  \BibitemOpen
  \bibfield  {author} {\bibinfo {author} {\bibfnamefont {C.~V.}\ \bibnamefont
  {Vishveshwara}},\ }\href {\doibase 10.1103/PhysRevD.1.2870} {\bibfield
  {journal} {\bibinfo  {journal} {Phys. Rev. D}\ }\textbf {\bibinfo {volume}
  {1}},\ \bibinfo {pages} {2870} (\bibinfo {year} {1970})}\BibitemShut
  {NoStop}%
\bibitem [{\citenamefont {Press}(1971)}]{Press1971}%
  \BibitemOpen
  \bibfield  {author} {\bibinfo {author} {\bibfnamefont {W.~H.}\ \bibnamefont
  {Press}},\ }\href {\doibase 10.1086/180849} {\bibfield  {journal} {\bibinfo
  {journal} {Astrophys. J.}\ }\textbf {\bibinfo {volume} {170}},\ \bibinfo
  {pages} {L105} (\bibinfo {year} {1971})}\BibitemShut {NoStop}%
\bibitem [{\citenamefont {{Teukolsky}}(1973)}]{Teukolsky}%
  \BibitemOpen
  \bibfield  {author} {\bibinfo {author} {\bibfnamefont {S.~A.}\ \bibnamefont
  {{Teukolsky}}},\ }\href {\doibase 10.1086/152444} {\bibfield  {journal}
  {\bibinfo  {journal} {\apj}\ }\textbf {\bibinfo {volume} {185}},\ \bibinfo
  {pages} {635} (\bibinfo {year} {1973})}\BibitemShut {NoStop}%
\bibitem [{\citenamefont {Chandrasekhar}\ and\ \citenamefont
  {Detweiler}(1975)}]{ChandraDetweiler1975}%
  \BibitemOpen
  \bibfield  {author} {\bibinfo {author} {\bibfnamefont {S.}~\bibnamefont
  {Chandrasekhar}}\ and\ \bibinfo {author} {\bibfnamefont {S.}~\bibnamefont
  {Detweiler}},\ }\href {http://www.jstor.org/stable/78902} {\bibfield
  {journal} {\bibinfo  {journal} {Proc.\ R.\ Soc.\ A}\ }\textbf {\bibinfo
  {volume} {344}},\ \bibinfo {pages} {441} (\bibinfo {year}
  {1975})}\BibitemShut {NoStop}%
\bibitem [{\citenamefont {Echeverria}(1989)}]{Echeverria:1989hg}%
  \BibitemOpen
  \bibfield  {author} {\bibinfo {author} {\bibfnamefont {F.}~\bibnamefont
  {Echeverria}},\ }\href {\doibase 10.1103/PhysRevD.40.3194} {\bibfield
  {journal} {\bibinfo  {journal} {Phys. Rev.}\ }\textbf {\bibinfo {volume}
  {D40}},\ \bibinfo {pages} {3194} (\bibinfo {year} {1989})}\BibitemShut
  {NoStop}%
\bibitem [{\citenamefont {Dreyer}\ \emph {et~al.}(2004)\citenamefont {Dreyer},
  \citenamefont {Kelly}, \citenamefont {Krishnan}, \citenamefont {Finn},
  \citenamefont {Garrison},\ and\ \citenamefont
  {Lopez-Aleman}}]{Dreyer:2003bv}%
  \BibitemOpen
  \bibfield  {author} {\bibinfo {author} {\bibfnamefont {O.}~\bibnamefont
  {Dreyer}}, \bibinfo {author} {\bibfnamefont {B.~J.}\ \bibnamefont {Kelly}},
  \bibinfo {author} {\bibfnamefont {B.}~\bibnamefont {Krishnan}}, \bibinfo
  {author} {\bibfnamefont {L.~S.}\ \bibnamefont {Finn}}, \bibinfo {author}
  {\bibfnamefont {D.}~\bibnamefont {Garrison}}, \ and\ \bibinfo {author}
  {\bibfnamefont {R.}~\bibnamefont {Lopez-Aleman}},\ }\href {\doibase
  10.1088/0264-9381/21/4/003} {\bibfield  {journal} {\bibinfo  {journal}
  {Class. Quant. Grav.}\ }\textbf {\bibinfo {volume} {21}},\ \bibinfo {pages}
  {787} (\bibinfo {year} {2004})},\ \Eprint
  {http://arxiv.org/abs/gr-qc/0309007} {arXiv:gr-qc/0309007 [gr-qc]}
  \BibitemShut {NoStop}%
\bibitem [{\citenamefont {Berti}\ \emph {et~al.}(2006)\citenamefont {Berti},
  \citenamefont {Cardoso},\ and\ \citenamefont {Will}}]{Berti:2005ys}%
  \BibitemOpen
  \bibfield  {author} {\bibinfo {author} {\bibfnamefont {E.}~\bibnamefont
  {Berti}}, \bibinfo {author} {\bibfnamefont {V.}~\bibnamefont {Cardoso}}, \
  and\ \bibinfo {author} {\bibfnamefont {C.~M.}\ \bibnamefont {Will}},\ }\href
  {\doibase 10.1103/PhysRevD.73.064030} {\bibfield  {journal} {\bibinfo
  {journal} {Phys. Rev.}\ }\textbf {\bibinfo {volume} {D73}},\ \bibinfo {pages}
  {064030} (\bibinfo {year} {2006})},\ \Eprint
  {http://arxiv.org/abs/gr-qc/0512160} {arXiv:gr-qc/0512160 [gr-qc]}
  \BibitemShut {NoStop}%
\bibitem [{\citenamefont {Gossan}\ \emph {et~al.}(2012)\citenamefont {Gossan},
  \citenamefont {Veitch},\ and\ \citenamefont {Sathyaprakash}}]{Gossan:2011ha}%
  \BibitemOpen
  \bibfield  {author} {\bibinfo {author} {\bibfnamefont {S.}~\bibnamefont
  {Gossan}}, \bibinfo {author} {\bibfnamefont {J.}~\bibnamefont {Veitch}}, \
  and\ \bibinfo {author} {\bibfnamefont {B.~S.}\ \bibnamefont
  {Sathyaprakash}},\ }\href {\doibase 10.1103/PhysRevD.85.124056} {\bibfield
  {journal} {\bibinfo  {journal} {Phys. Rev.}\ }\textbf {\bibinfo {volume}
  {D85}},\ \bibinfo {pages} {124056} (\bibinfo {year} {2012})},\ \Eprint
  {http://arxiv.org/abs/1111.5819} {arXiv:1111.5819 [gr-qc]} \BibitemShut
  {NoStop}%
\bibitem [{\citenamefont {Meidam}\ \emph {et~al.}(2014)\citenamefont {Meidam},
  \citenamefont {Agathos}, \citenamefont {Van Den~Broeck}, \citenamefont
  {Veitch},\ and\ \citenamefont {Sathyaprakash}}]{Meidam:2014jpa}%
  \BibitemOpen
  \bibfield  {author} {\bibinfo {author} {\bibfnamefont {J.}~\bibnamefont
  {Meidam}}, \bibinfo {author} {\bibfnamefont {M.}~\bibnamefont {Agathos}},
  \bibinfo {author} {\bibfnamefont {C.}~\bibnamefont {Van Den~Broeck}},
  \bibinfo {author} {\bibfnamefont {J.}~\bibnamefont {Veitch}}, \ and\ \bibinfo
  {author} {\bibfnamefont {B.~S.}\ \bibnamefont {Sathyaprakash}},\ }\href
  {\doibase 10.1103/PhysRevD.90.064009} {\bibfield  {journal} {\bibinfo
  {journal} {Phys. Rev.}\ }\textbf {\bibinfo {volume} {D90}},\ \bibinfo {pages}
  {064009} (\bibinfo {year} {2014})},\ \Eprint {http://arxiv.org/abs/1406.3201}
  {arXiv:1406.3201 [gr-qc]} \BibitemShut {NoStop}%
\bibitem [{\citenamefont {Berti}\ \emph {et~al.}(2015)\citenamefont {Berti}
  \emph {et~al.}}]{Berti:2015itd}%
  \BibitemOpen
  \bibfield  {author} {\bibinfo {author} {\bibfnamefont {E.}~\bibnamefont
  {Berti}} \emph {et~al.},\ }\href {\doibase 10.1088/0264-9381/32/24/243001}
  {\bibfield  {journal} {\bibinfo  {journal} {Class. Quant. Grav.}\ }\textbf
  {\bibinfo {volume} {32}},\ \bibinfo {pages} {243001} (\bibinfo {year}
  {2015})},\ \Eprint {http://arxiv.org/abs/1501.07274} {arXiv:1501.07274
  [gr-qc]} \BibitemShut {NoStop}%
\bibitem [{\citenamefont {Berti}\ \emph {et~al.}(2016)\citenamefont {Berti},
  \citenamefont {Sesana}, \citenamefont {Barausse}, \citenamefont {Cardoso},\
  and\ \citenamefont {Belczynski}}]{Berti:2016lat}%
  \BibitemOpen
  \bibfield  {author} {\bibinfo {author} {\bibfnamefont {E.}~\bibnamefont
  {Berti}}, \bibinfo {author} {\bibfnamefont {A.}~\bibnamefont {Sesana}},
  \bibinfo {author} {\bibfnamefont {E.}~\bibnamefont {Barausse}}, \bibinfo
  {author} {\bibfnamefont {V.}~\bibnamefont {Cardoso}}, \ and\ \bibinfo
  {author} {\bibfnamefont {K.}~\bibnamefont {Belczynski}},\ }\href {\doibase
  10.1103/PhysRevLett.117.101102} {\bibfield  {journal} {\bibinfo  {journal}
  {Phys. Rev. Lett.}\ }\textbf {\bibinfo {volume} {117}},\ \bibinfo {pages}
  {101102} (\bibinfo {year} {2016})},\ \Eprint
  {http://arxiv.org/abs/1605.09286} {arXiv:1605.09286 [gr-qc]} \BibitemShut
  {NoStop}%
\bibitem [{\citenamefont {Baibhav}\ \emph {et~al.}(2018)\citenamefont
  {Baibhav}, \citenamefont {Berti}, \citenamefont {Cardoso},\ and\
  \citenamefont {Khanna}}]{Baibhav:2017jhs}%
  \BibitemOpen
  \bibfield  {author} {\bibinfo {author} {\bibfnamefont {V.}~\bibnamefont
  {Baibhav}}, \bibinfo {author} {\bibfnamefont {E.}~\bibnamefont {Berti}},
  \bibinfo {author} {\bibfnamefont {V.}~\bibnamefont {Cardoso}}, \ and\
  \bibinfo {author} {\bibfnamefont {G.}~\bibnamefont {Khanna}},\ }\href
  {\doibase 10.1103/PhysRevD.97.044048} {\bibfield  {journal} {\bibinfo
  {journal} {Phys. Rev.}\ }\textbf {\bibinfo {volume} {D97}},\ \bibinfo {pages}
  {044048} (\bibinfo {year} {2018})},\ \Eprint
  {http://arxiv.org/abs/1710.02156} {arXiv:1710.02156 [gr-qc]} \BibitemShut
  {NoStop}%
\bibitem [{\citenamefont {Baibhav}\ and\ \citenamefont
  {Berti}(2019)}]{Baibhav:2018rfk}%
  \BibitemOpen
  \bibfield  {author} {\bibinfo {author} {\bibfnamefont {V.}~\bibnamefont
  {Baibhav}}\ and\ \bibinfo {author} {\bibfnamefont {E.}~\bibnamefont
  {Berti}},\ }\href {\doibase 10.1103/PhysRevD.99.024005} {\bibfield  {journal}
  {\bibinfo  {journal} {Phys. Rev.}\ }\textbf {\bibinfo {volume} {D99}},\
  \bibinfo {pages} {024005} (\bibinfo {year} {2019})},\ \Eprint
  {http://arxiv.org/abs/1809.03500} {arXiv:1809.03500 [gr-qc]} \BibitemShut
  {NoStop}%
\bibitem [{\citenamefont {Aasi}\ \emph {et~al.}(2015)\citenamefont {Aasi} \emph
  {et~al.}}]{aLIGO}%
  \BibitemOpen
  \bibfield  {author} {\bibinfo {author} {\bibfnamefont {J.}~\bibnamefont
  {Aasi}} \emph {et~al.},\ }\href {\doibase 10.1088/0264-9381/32/7/074001}
  {\bibfield  {journal} {\bibinfo  {journal} {Class. Quant. Grav.}\ }\textbf
  {\bibinfo {volume} {32}},\ \bibinfo {pages} {074001} (\bibinfo {year}
  {2015})}\BibitemShut {NoStop}%
\bibitem [{\citenamefont {Acernese}\ \emph {et~al.}(2015)\citenamefont
  {Acernese} \emph {et~al.}}]{Virgo}%
  \BibitemOpen
  \bibfield  {author} {\bibinfo {author} {\bibfnamefont {F.}~\bibnamefont
  {Acernese}} \emph {et~al.},\ }\href {\doibase 10.1088/0264-9381/32/2/024001}
  {\bibfield  {journal} {\bibinfo  {journal} {Class. Quant. Grav.}\ }\textbf
  {\bibinfo {volume} {32}},\ \bibinfo {pages} {024001} (\bibinfo {year}
  {2015})}\BibitemShut {NoStop}%
\bibitem [{\citenamefont {Abbott}\ \emph
  {et~al.}(2016{\natexlab{a}})\citenamefont {Abbott} \emph
  {et~al.}}]{gw150914}%
  \BibitemOpen
  \bibfield  {author} {\bibinfo {author} {\bibfnamefont {B.~P.}\ \bibnamefont
  {Abbott}} \emph {et~al.} (\bibinfo {collaboration} {LIGO Scientific
  Collaboration, Virgo Collaboration}),\ }\href {\doibase
  10.1103/PhysRevLett.116.061102} {\bibfield  {journal} {\bibinfo  {journal}
  {Phys. Rev. Lett.}\ }\textbf {\bibinfo {volume} {116}},\ \bibinfo {pages}
  {061102} (\bibinfo {year} {2016}{\natexlab{a}})}\BibitemShut {NoStop}%
\bibitem [{\citenamefont {Abbott}\ \emph
  {et~al.}(2016{\natexlab{b}})\citenamefont {Abbott} \emph
  {et~al.}}]{gw151226}%
  \BibitemOpen
  \bibfield  {author} {\bibinfo {author} {\bibfnamefont {B.~P.}\ \bibnamefont
  {Abbott}} \emph {et~al.} (\bibinfo {collaboration} {LIGO Scientific
  Collaboration, Virgo Collaboration}),\ }\href {\doibase
  10.1103/PhysRevLett.116.241103} {\bibfield  {journal} {\bibinfo  {journal}
  {Phys. Rev. Lett.}\ }\textbf {\bibinfo {volume} {116}},\ \bibinfo {pages}
  {241103} (\bibinfo {year} {2016}{\natexlab{b}})}\BibitemShut {NoStop}%
\bibitem [{\citenamefont {Abbott}\ \emph
  {et~al.}(2016{\natexlab{c}})\citenamefont {Abbott} \emph {et~al.}}]{o1bbh}%
  \BibitemOpen
  \bibfield  {author} {\bibinfo {author} {\bibfnamefont {B.~P.}\ \bibnamefont
  {Abbott}} \emph {et~al.} (\bibinfo {collaboration} {LIGO Scientific
  Collaboration, Virgo Collaboration}),\ }\href {\doibase
  10.1103/PhysRevX.6.041015} {\bibfield  {journal} {\bibinfo  {journal} {Phys.
  Rev. X}\ }\textbf {\bibinfo {volume} {6}},\ \bibinfo {pages} {041015}
  (\bibinfo {year} {2016}{\natexlab{c}})}\BibitemShut {NoStop}%
\bibitem [{\citenamefont {Abbott}\ \emph
  {et~al.}(2017{\natexlab{a}})\citenamefont {Abbott} \emph
  {et~al.}}]{gw170104}%
  \BibitemOpen
  \bibfield  {author} {\bibinfo {author} {\bibfnamefont {B.~P.}\ \bibnamefont
  {Abbott}} \emph {et~al.} (\bibinfo {collaboration} {LIGO Scientific
  Collaboration, Virgo Collaboration}),\ }\href {\doibase
  10.1103/PhysRevLett.118.221101} {\bibfield  {journal} {\bibinfo  {journal}
  {Phys. Rev. Lett.}\ }\textbf {\bibinfo {volume} {118}},\ \bibinfo {pages}
  {221101} (\bibinfo {year} {2017}{\natexlab{a}})}\BibitemShut {NoStop}%
\bibitem [{\citenamefont {Abbott}\ \emph
  {et~al.}(2017{\natexlab{b}})\citenamefont {Abbott} \emph
  {et~al.}}]{gw170608}%
  \BibitemOpen
  \bibfield  {author} {\bibinfo {author} {\bibfnamefont {B.~P.}\ \bibnamefont
  {Abbott}} \emph {et~al.} (\bibinfo {collaboration} {LIGO Scientific
  Collaboration, Virgo Collaboration}),\ }\href {\doibase
  10.3847/2041-8213/aa9f0c} {\bibfield  {journal} {\bibinfo  {journal}
  {Astrophys. J.}\ }\textbf {\bibinfo {volume} {851}},\ \bibinfo {pages} {L35}
  (\bibinfo {year} {2017}{\natexlab{b}})}\BibitemShut {NoStop}%
\bibitem [{\citenamefont {Abbott}\ \emph
  {et~al.}(2017{\natexlab{c}})\citenamefont {Abbott} \emph
  {et~al.}}]{gw170814}%
  \BibitemOpen
  \bibfield  {author} {\bibinfo {author} {\bibfnamefont {B.~P.}\ \bibnamefont
  {Abbott}} \emph {et~al.} (\bibinfo {collaboration} {LIGO Scientific
  Collaboration, Virgo Collaboration}),\ }\href {\doibase
  10.1103/PhysRevLett.119.141101} {\bibfield  {journal} {\bibinfo  {journal}
  {Phys. Rev. Lett.}\ }\textbf {\bibinfo {volume} {119}},\ \bibinfo {pages}
  {141101} (\bibinfo {year} {2017}{\natexlab{c}})}\BibitemShut {NoStop}%
\bibitem [{\citenamefont {Abbott}\ \emph {et~al.}(2018)\citenamefont {Abbott}
  \emph {et~al.}}]{gwtc1:2018}%
  \BibitemOpen
  \bibfield  {author} {\bibinfo {author} {\bibfnamefont {B.~P.}\ \bibnamefont
  {Abbott}} \emph {et~al.} (\bibinfo {collaboration} {LIGO Scientific,
  Virgo}),\ }\href@noop {} {\  (\bibinfo {year} {2018})},\ \Eprint
  {http://arxiv.org/abs/1811.12907} {arXiv:1811.12907 [astro-ph.HE]}
  \BibitemShut {NoStop}%
\bibitem [{\citenamefont {Abbott}\ \emph
  {et~al.}(2016{\natexlab{d}})\citenamefont {Abbott} \emph
  {et~al.}}]{gw150914_tgr}%
  \BibitemOpen
  \bibfield  {author} {\bibinfo {author} {\bibfnamefont {B.~P.}\ \bibnamefont
  {Abbott}} \emph {et~al.} (\bibinfo {collaboration} {LIGO Scientific
  Collaboration, Virgo Collaboration}),\ }\href {\doibase
  10.1103/PhysRevLett.116.221101} {\bibfield  {journal} {\bibinfo  {journal}
  {Phys. Rev. Lett.}\ }\textbf {\bibinfo {volume} {116}},\ \bibinfo {pages}
  {221101} (\bibinfo {year} {2016}{\natexlab{d}})}\BibitemShut {NoStop}%
\bibitem [{\citenamefont {Del~Pozzo}\ and\ \citenamefont
  {Nagar}(2017)}]{Nagar:2016iwa}%
  \BibitemOpen
  \bibfield  {author} {\bibinfo {author} {\bibfnamefont {W.}~\bibnamefont
  {Del~Pozzo}}\ and\ \bibinfo {author} {\bibfnamefont {A.}~\bibnamefont
  {Nagar}},\ }\href {\doibase 10.1103/PhysRevD.95.124034} {\bibfield  {journal}
  {\bibinfo  {journal} {Phys. Rev.}\ }\textbf {\bibinfo {volume} {D95}},\
  \bibinfo {pages} {124034} (\bibinfo {year} {2017})},\ \Eprint
  {http://arxiv.org/abs/1606.03952} {arXiv:1606.03952 [gr-qc]} \BibitemShut
  {NoStop}%
\bibitem [{\citenamefont {Cabero}\ \emph {et~al.}(2018)\citenamefont {Cabero},
  \citenamefont {Capano}, \citenamefont {Fischer-Birnholtz}, \citenamefont
  {Krishnan}, \citenamefont {Nielsen}, \citenamefont {Nitz},\ and\
  \citenamefont {Biwer}}]{Cabero:2017avf}%
  \BibitemOpen
  \bibfield  {author} {\bibinfo {author} {\bibfnamefont {M.}~\bibnamefont
  {Cabero}}, \bibinfo {author} {\bibfnamefont {C.~D.}\ \bibnamefont {Capano}},
  \bibinfo {author} {\bibfnamefont {O.}~\bibnamefont {Fischer-Birnholtz}},
  \bibinfo {author} {\bibfnamefont {B.}~\bibnamefont {Krishnan}}, \bibinfo
  {author} {\bibfnamefont {A.~B.}\ \bibnamefont {Nielsen}}, \bibinfo {author}
  {\bibfnamefont {A.~H.}\ \bibnamefont {Nitz}}, \ and\ \bibinfo {author}
  {\bibfnamefont {C.~M.}\ \bibnamefont {Biwer}},\ }\href {\doibase
  10.1103/PhysRevD.97.124069} {\bibfield  {journal} {\bibinfo  {journal} {Phys.
  Rev.}\ }\textbf {\bibinfo {volume} {D97}},\ \bibinfo {pages} {124069}
  (\bibinfo {year} {2018})},\ \Eprint {http://arxiv.org/abs/1711.09073}
  {arXiv:1711.09073 [gr-qc]} \BibitemShut {NoStop}%
\bibitem [{\citenamefont {Thrane}\ \emph {et~al.}(2017)\citenamefont {Thrane},
  \citenamefont {Lasky},\ and\ \citenamefont {Levin}}]{Thrane:2017lqn}%
  \BibitemOpen
  \bibfield  {author} {\bibinfo {author} {\bibfnamefont {E.}~\bibnamefont
  {Thrane}}, \bibinfo {author} {\bibfnamefont {P.~D.}\ \bibnamefont {Lasky}}, \
  and\ \bibinfo {author} {\bibfnamefont {Y.}~\bibnamefont {Levin}},\ }\href
  {\doibase 10.1103/PhysRevD.96.102004} {\bibfield  {journal} {\bibinfo
  {journal} {Phys. Rev.}\ }\textbf {\bibinfo {volume} {D96}},\ \bibinfo {pages}
  {102004} (\bibinfo {year} {2017})},\ \Eprint
  {http://arxiv.org/abs/1706.05152} {arXiv:1706.05152 [gr-qc]} \BibitemShut
  {NoStop}%
\bibitem [{\citenamefont {Brito}\ \emph {et~al.}(2018)\citenamefont {Brito},
  \citenamefont {Buonanno},\ and\ \citenamefont {Raymond}}]{Brito:2018rfr}%
  \BibitemOpen
  \bibfield  {author} {\bibinfo {author} {\bibfnamefont {R.}~\bibnamefont
  {Brito}}, \bibinfo {author} {\bibfnamefont {A.}~\bibnamefont {Buonanno}}, \
  and\ \bibinfo {author} {\bibfnamefont {V.}~\bibnamefont {Raymond}},\ }\href
  {\doibase 10.1103/PhysRevD.98.084038} {\bibfield  {journal} {\bibinfo
  {journal} {Phys. Rev.}\ }\textbf {\bibinfo {volume} {D98}},\ \bibinfo {pages}
  {084038} (\bibinfo {year} {2018})},\ \Eprint
  {http://arxiv.org/abs/1805.00293} {arXiv:1805.00293 [gr-qc]} \BibitemShut
  {NoStop}%
\bibitem [{\citenamefont {Carullo}\ \emph {et~al.}(2018)\citenamefont {Carullo}
  \emph {et~al.}}]{Carullo:2018sfu}%
  \BibitemOpen
  \bibfield  {author} {\bibinfo {author} {\bibfnamefont {G.}~\bibnamefont
  {Carullo}} \emph {et~al.},\ }\href {\doibase 10.1103/PhysRevD.98.104020}
  {\bibfield  {journal} {\bibinfo  {journal} {Phys. Rev.}\ }\textbf {\bibinfo
  {volume} {D98}},\ \bibinfo {pages} {104020} (\bibinfo {year} {2018})},\
  \Eprint {http://arxiv.org/abs/1805.04760} {arXiv:1805.04760 [gr-qc]}
  \BibitemShut {NoStop}%
\bibitem [{\citenamefont {Carullo}\ \emph {et~al.}(2019)\citenamefont
  {Carullo}, \citenamefont {Del~Pozzo},\ and\ \citenamefont
  {Veitch}}]{Carullo:2019flw}%
  \BibitemOpen
  \bibfield  {author} {\bibinfo {author} {\bibfnamefont {G.}~\bibnamefont
  {Carullo}}, \bibinfo {author} {\bibfnamefont {W.}~\bibnamefont {Del~Pozzo}},
  \ and\ \bibinfo {author} {\bibfnamefont {J.}~\bibnamefont {Veitch}},\ }\href
  {\doibase 10.1103/PhysRevD.99.123029} {\bibfield  {journal} {\bibinfo
  {journal} {Phys. Rev.}\ }\textbf {\bibinfo {volume} {D99}},\ \bibinfo {pages}
  {123029} (\bibinfo {year} {2019})},\ \Eprint
  {http://arxiv.org/abs/1902.07527} {arXiv:1902.07527 [gr-qc]} \BibitemShut
  {NoStop}%
\bibitem [{\citenamefont {Kamaretsos}\ \emph {et~al.}(2012)\citenamefont
  {Kamaretsos}, \citenamefont {Hannam}, \citenamefont {Husa},\ and\
  \citenamefont {Sathyaprakash}}]{Kamaretsos:2011um}%
  \BibitemOpen
  \bibfield  {author} {\bibinfo {author} {\bibfnamefont {I.}~\bibnamefont
  {Kamaretsos}}, \bibinfo {author} {\bibfnamefont {M.}~\bibnamefont {Hannam}},
  \bibinfo {author} {\bibfnamefont {S.}~\bibnamefont {Husa}}, \ and\ \bibinfo
  {author} {\bibfnamefont {B.~S.}\ \bibnamefont {Sathyaprakash}},\ }\href
  {\doibase 10.1103/PhysRevD.85.024018} {\bibfield  {journal} {\bibinfo
  {journal} {Phys. Rev.}\ }\textbf {\bibinfo {volume} {D85}},\ \bibinfo {pages}
  {024018} (\bibinfo {year} {2012})},\ \Eprint {http://arxiv.org/abs/1107.0854}
  {arXiv:1107.0854 [gr-qc]} \BibitemShut {NoStop}%
\bibitem [{\citenamefont {London}\ \emph {et~al.}(2014)\citenamefont {London},
  \citenamefont {Healy},\ and\ \citenamefont {Shoemaker}}]{London:2014cma}%
  \BibitemOpen
  \bibfield  {author} {\bibinfo {author} {\bibfnamefont {L.}~\bibnamefont
  {London}}, \bibinfo {author} {\bibfnamefont {J.}~\bibnamefont {Healy}}, \
  and\ \bibinfo {author} {\bibfnamefont {D.}~\bibnamefont {Shoemaker}},\
  }\href@noop {} {\bibfield  {journal} {\bibinfo  {journal} {Phys.\ Rev.\ D}\
  }\textbf {\bibinfo {volume} {90}},\ \bibinfo {pages} {124032} (\bibinfo
  {year} {2014})},\ \Eprint {http://arxiv.org/abs/1404.3197} {arXiv:1404.3197
  [gr-qc]} \BibitemShut {NoStop}%
\bibitem [{\citenamefont {Giesler}\ \emph {et~al.}(2019)\citenamefont
  {Giesler}, \citenamefont {Isi}, \citenamefont {Scheel},\ and\ \citenamefont
  {Teukolsky}}]{Giesler:2019uxc}%
  \BibitemOpen
  \bibfield  {author} {\bibinfo {author} {\bibfnamefont {M.}~\bibnamefont
  {Giesler}}, \bibinfo {author} {\bibfnamefont {M.}~\bibnamefont {Isi}},
  \bibinfo {author} {\bibfnamefont {M.}~\bibnamefont {Scheel}}, \ and\ \bibinfo
  {author} {\bibfnamefont {S.}~\bibnamefont {Teukolsky}},\ }\href@noop {}
  {\enquote {\bibinfo {title} {{Black hole ringdown: the importance of
  overtones}},}\ } (\bibinfo {year} {2019}),\ \Eprint
  {http://arxiv.org/abs/1903.08284} {arXiv:1903.08284 [gr-qc]} \BibitemShut
  {NoStop}%
\bibitem [{\citenamefont {Buonanno}\ \emph {et~al.}(2007)\citenamefont
  {Buonanno}, \citenamefont {Cook},\ and\ \citenamefont
  {Pretorius}}]{Buonanno:2006ui}%
  \BibitemOpen
  \bibfield  {author} {\bibinfo {author} {\bibfnamefont {A.}~\bibnamefont
  {Buonanno}}, \bibinfo {author} {\bibfnamefont {G.~B.}\ \bibnamefont {Cook}},
  \ and\ \bibinfo {author} {\bibfnamefont {F.}~\bibnamefont {Pretorius}},\
  }\href {\doibase 10.1103/PhysRevD.75.124018} {\bibfield  {journal} {\bibinfo
  {journal} {Phys. Rev.}\ }\textbf {\bibinfo {volume} {D75}},\ \bibinfo {pages}
  {124018} (\bibinfo {year} {2007})},\ \Eprint
  {http://arxiv.org/abs/gr-qc/0610122} {arXiv:gr-qc/0610122 [gr-qc]}
  \BibitemShut {NoStop}%
\bibitem [{\citenamefont {Pan}\ \emph {et~al.}(2013)\citenamefont {Pan},
  \citenamefont {Buonanno}, \citenamefont {Taracchini}, \citenamefont {Kidder},
  \citenamefont {Mroue} \emph {et~al.}}]{Pan:2013rra}%
  \BibitemOpen
  \bibfield  {author} {\bibinfo {author} {\bibfnamefont {Y.}~\bibnamefont
  {Pan}}, \bibinfo {author} {\bibfnamefont {A.}~\bibnamefont {Buonanno}},
  \bibinfo {author} {\bibfnamefont {A.}~\bibnamefont {Taracchini}}, \bibinfo
  {author} {\bibfnamefont {L.~E.}\ \bibnamefont {Kidder}}, \bibinfo {author}
  {\bibfnamefont {A.~H.}\ \bibnamefont {Mroue}},  \emph {et~al.},\ }\href@noop
  {} {\bibfield  {journal} {\bibinfo  {journal} {Phys.\ Rev.\ D}\ }\textbf
  {\bibinfo {volume} {89}},\ \bibinfo {pages} {084006} (\bibinfo {year}
  {2013})},\ \Eprint {http://arxiv.org/abs/1307.6232} {arXiv:1307.6232 [gr-qc]}
  \BibitemShut {NoStop}%
\bibitem [{\citenamefont {Taracchini}\ \emph {et~al.}(2014)\citenamefont
  {Taracchini}, \citenamefont {Buonanno}, \citenamefont {Pan}, \citenamefont
  {Hinderer}, \citenamefont {Boyle}, \citenamefont {Hemberger}, \citenamefont
  {Kidder}, \citenamefont {Lovelace}, \citenamefont {Mroue}, \citenamefont
  {Pfeiffer}, \citenamefont {Scheel}, \citenamefont {Szilagyi}, \citenamefont
  {Taylor},\ and\ \citenamefont {Zenginoglu}}]{Taracchini:2013rva}%
  \BibitemOpen
  \bibfield  {author} {\bibinfo {author} {\bibfnamefont {A.}~\bibnamefont
  {Taracchini}}, \bibinfo {author} {\bibfnamefont {A.}~\bibnamefont
  {Buonanno}}, \bibinfo {author} {\bibfnamefont {Y.}~\bibnamefont {Pan}},
  \bibinfo {author} {\bibfnamefont {T.}~\bibnamefont {Hinderer}}, \bibinfo
  {author} {\bibfnamefont {M.}~\bibnamefont {Boyle}}, \bibinfo {author}
  {\bibfnamefont {D.~A.}\ \bibnamefont {Hemberger}}, \bibinfo {author}
  {\bibfnamefont {L.~E.}\ \bibnamefont {Kidder}}, \bibinfo {author}
  {\bibfnamefont {G.}~\bibnamefont {Lovelace}}, \bibinfo {author}
  {\bibfnamefont {A.~H.}\ \bibnamefont {Mroue}}, \bibinfo {author}
  {\bibfnamefont {H.~P.}\ \bibnamefont {Pfeiffer}}, \bibinfo {author}
  {\bibfnamefont {M.~A.}\ \bibnamefont {Scheel}}, \bibinfo {author}
  {\bibfnamefont {B.}~\bibnamefont {Szilagyi}}, \bibinfo {author}
  {\bibfnamefont {N.~W.}\ \bibnamefont {Taylor}}, \ and\ \bibinfo {author}
  {\bibfnamefont {A.}~\bibnamefont {Zenginoglu}},\ }\href {\doibase
  10.1103/PhysRevD.89.061502} {\bibfield  {journal} {\bibinfo  {journal}
  {Phys.\ Rev.\ D}\ }\textbf {\bibinfo {volume} {89 (R)}},\ \bibinfo {pages}
  {061502} (\bibinfo {year} {2014})},\ \Eprint {http://arxiv.org/abs/1311.2544}
  {arXiv:1311.2544 [gr-qc]} \BibitemShut {NoStop}%
\bibitem [{\citenamefont {Babak}\ \emph {et~al.}(2016)\citenamefont {Babak},
  \citenamefont {Taracchini},\ and\ \citenamefont {Buonanno}}]{Babak:2016tgq}%
  \BibitemOpen
  \bibfield  {author} {\bibinfo {author} {\bibfnamefont {S.}~\bibnamefont
  {Babak}}, \bibinfo {author} {\bibfnamefont {A.}~\bibnamefont {Taracchini}}, \
  and\ \bibinfo {author} {\bibfnamefont {A.}~\bibnamefont {Buonanno}},\
  }\href@noop {} {\  (\bibinfo {year} {2016})},\ \Eprint
  {http://arxiv.org/abs/1607.05661} {arXiv:1607.05661 [gr-qc]} \BibitemShut
  {NoStop}%
\bibitem [{\citenamefont {Bohe}\ \emph {et~al.}(2017)\citenamefont {Bohe} \emph
  {et~al.}}]{Bohe:2016gbl}%
  \BibitemOpen
  \bibfield  {author} {\bibinfo {author} {\bibfnamefont {A.}~\bibnamefont
  {Bohe}} \emph {et~al.},\ }\href {\doibase 10.1103/PhysRevD.95.044028}
  {\bibfield  {journal} {\bibinfo  {journal} {Phys. Rev.}\ }\textbf {\bibinfo
  {volume} {D95}},\ \bibinfo {pages} {044028} (\bibinfo {year} {2017})},\
  \Eprint {http://arxiv.org/abs/1611.03703} {arXiv:1611.03703 [gr-qc]}
  \BibitemShut {NoStop}%
\bibitem [{\citenamefont {Bhagwat}\ \emph {et~al.}(2016)\citenamefont
  {Bhagwat}, \citenamefont {Brown},\ and\ \citenamefont
  {Ballmer}}]{Bhagwat:2016ntk}%
  \BibitemOpen
  \bibfield  {author} {\bibinfo {author} {\bibfnamefont {S.}~\bibnamefont
  {Bhagwat}}, \bibinfo {author} {\bibfnamefont {D.~A.}\ \bibnamefont {Brown}},
  \ and\ \bibinfo {author} {\bibfnamefont {S.~W.}\ \bibnamefont {Ballmer}},\
  }\href {\doibase 10.1103/PhysRevD.94.084024, 10.1103/PhysRevD.95.069906}
  {\bibfield  {journal} {\bibinfo  {journal} {Phys. Rev.}\ }\textbf {\bibinfo
  {volume} {D94}},\ \bibinfo {pages} {084024} (\bibinfo {year} {2016})},\
  \bibinfo {note} {[Erratum: Phys. Rev.D95,no.6,069906(2017)]},\ \Eprint
  {http://arxiv.org/abs/1607.07845} {arXiv:1607.07845 [gr-qc]} \BibitemShut
  {NoStop}%
\bibitem [{\citenamefont {Abbott}\ \emph
  {et~al.}(2016{\natexlab{e}})\citenamefont {Abbott} \emph
  {et~al.}}]{TheLIGOScientific:2016src}%
  \BibitemOpen
  \bibfield  {author} {\bibinfo {author} {\bibfnamefont {B.~P.}\ \bibnamefont
  {Abbott}} \emph {et~al.} (\bibinfo {collaboration} {LIGO Scientific
  Collaboration, Virgo Collaboration}),\ }\href@noop {} {\bibfield  {journal}
  {\bibinfo  {journal} {Phys.~Rev.~Lett.}\ }\textbf {\bibinfo {volume} {116}},\
  \bibinfo {pages} {221101} (\bibinfo {year} {2016}{\natexlab{e}})},\ \Eprint
  {http://arxiv.org/abs/1602.03841} {arXiv:1602.03841 [gr-qc]} \BibitemShut
  {NoStop}%
\bibitem [{\citenamefont {Teukolsky}(1972)}]{Teukolsky:1972my}%
  \BibitemOpen
  \bibfield  {author} {\bibinfo {author} {\bibfnamefont {S.}~\bibnamefont
  {Teukolsky}},\ }\href {\doibase 10.1103/PhysRevLett.29.1114} {\bibfield
  {journal} {\bibinfo  {journal} {Phys.\ Rev.\ Lett.}\ }\textbf {\bibinfo
  {volume} {29}},\ \bibinfo {pages} {1114} (\bibinfo {year}
  {1972})}\BibitemShut {NoStop}%
\bibitem [{\citenamefont {{Press}}\ and\ \citenamefont
  {{Teukolsky}}(1973)}]{Press:1973}%
  \BibitemOpen
  \bibfield  {author} {\bibinfo {author} {\bibfnamefont {W.~H.}\ \bibnamefont
  {{Press}}}\ and\ \bibinfo {author} {\bibfnamefont {S.~A.}\ \bibnamefont
  {{Teukolsky}}},\ }\href {\doibase 10.1086/152445} {\bibfield  {journal}
  {\bibinfo  {journal} {\apj}\ }\textbf {\bibinfo {volume} {185}},\ \bibinfo
  {pages} {649} (\bibinfo {year} {1973})}\BibitemShut {NoStop}%
\bibitem [{\citenamefont {Berti}\ and\ \citenamefont
  {Klein}(2014)}]{Berti:2014fga}%
  \BibitemOpen
  \bibfield  {author} {\bibinfo {author} {\bibfnamefont {E.}~\bibnamefont
  {Berti}}\ and\ \bibinfo {author} {\bibfnamefont {A.}~\bibnamefont {Klein}},\
  }\href {\doibase 10.1103/PhysRevD.90.064012} {\bibfield  {journal} {\bibinfo
  {journal} {Phys. Rev.}\ }\textbf {\bibinfo {volume} {D90}},\ \bibinfo {pages}
  {064012} (\bibinfo {year} {2014})},\ \Eprint {http://arxiv.org/abs/1408.1860}
  {arXiv:1408.1860 [gr-qc]} \BibitemShut {NoStop}%
\bibitem [{\citenamefont {Abbott}\ \emph
  {et~al.}(2016{\natexlab{f}})\citenamefont {Abbott} \emph
  {et~al.}}]{Abbott:2016apu}%
  \BibitemOpen
  \bibfield  {author} {\bibinfo {author} {\bibfnamefont {B.~P.}\ \bibnamefont
  {Abbott}} \emph {et~al.} (\bibinfo {collaboration} {LIGO Scientific
  Collaboration, Virgo Collaboration}),\ }\href@noop {} {\bibfield  {journal}
  {\bibinfo  {journal} {Phys.~Rev.~D}\ }\textbf {\bibinfo {volume} {94}},\
  \bibinfo {pages} {064035} (\bibinfo {year} {2016}{\natexlab{f}})},\ \Eprint
  {http://arxiv.org/abs/1606.01262} {arXiv:1606.01262 [gr-qc]} \BibitemShut
  {NoStop}%
\bibitem [{\citenamefont {Abbott}\ \emph
  {et~al.}(2017{\natexlab{d}})\citenamefont {Abbott} \emph
  {et~al.}}]{Abbott:2016wiq}%
  \BibitemOpen
  \bibfield  {author} {\bibinfo {author} {\bibfnamefont {B.~P.}\ \bibnamefont
  {Abbott}} \emph {et~al.} (\bibinfo {collaboration} {LIGO Scientific
  Collaboration, Virgo Collaboration}),\ }\href {\doibase
  10.1088/1361-6382/aa6854} {\bibfield  {journal} {\bibinfo  {journal} {Class.
  Quant. Grav.}\ }\textbf {\bibinfo {volume} {34}},\ \bibinfo {pages} {104002}
  (\bibinfo {year} {2017}{\natexlab{d}})},\ \Eprint
  {http://arxiv.org/abs/1611.07531} {arXiv:1611.07531 [gr-qc]} \BibitemShut
  {NoStop}%
\bibitem [{\citenamefont {Leaver}(1985)}]{Leaver1985}%
  \BibitemOpen
  \bibfield  {author} {\bibinfo {author} {\bibfnamefont {E.~W.}\ \bibnamefont
  {Leaver}},\ }\href {\doibase 10.1098/rspa.1985.0119} {\bibfield  {journal}
  {\bibinfo  {journal} {Proc. Roy. Soc. Lond.}\ }\textbf {\bibinfo {volume}
  {A402}},\ \bibinfo {pages} {285} (\bibinfo {year} {1985})}\BibitemShut
  {NoStop}%
\bibitem [{\citenamefont {{Berti}}\ \emph {et~al.}(2009)\citenamefont
  {{Berti}}, \citenamefont {{Cardoso}},\ and\ \citenamefont
  {{Starinets}}}]{Berti2009}%
  \BibitemOpen
  \bibfield  {author} {\bibinfo {author} {\bibfnamefont {E.}~\bibnamefont
  {{Berti}}}, \bibinfo {author} {\bibfnamefont {V.}~\bibnamefont {{Cardoso}}},
  \ and\ \bibinfo {author} {\bibfnamefont {A.~O.}\ \bibnamefont
  {{Starinets}}},\ }\href@noop {} {\bibfield  {journal} {\bibinfo  {journal}
  {Class.\ Quantum Grav.}\ }\textbf {\bibinfo {volume} {26}},\ \bibinfo {pages}
  {163001} (\bibinfo {year} {2009})},\ \Eprint {http://arxiv.org/abs/0905.2975}
  {arXiv:0905.2975 [gr-qc]} \BibitemShut {NoStop}%
\bibitem [{Ber()}]{BertiWebsite}%
  \BibitemOpen
  \href@noop {} {}\bibinfo {howpublished}
  {\url{http://pages.jh.edu/~eberti2/ringdown}}\BibitemShut {NoStop}%
\bibitem [{gwo()}]{gwosc}%
  \BibitemOpen
  \href@noop {} {}\bibinfo {howpublished}
  {\url{https://www.gw-openscience.org}}\BibitemShut {NoStop}%
\bibitem [{\citenamefont {Abbott}\ \emph
  {et~al.}(2016{\natexlab{g}})\citenamefont {Abbott} \emph
  {et~al.}}]{gw150914_pe}%
  \BibitemOpen
  \bibfield  {author} {\bibinfo {author} {\bibfnamefont {B.~P.}\ \bibnamefont
  {Abbott}} \emph {et~al.} (\bibinfo {collaboration} {LIGO Scientific
  Collaboration, Virgo Collaboration}),\ }\href {\doibase
  10.1103/PhysRevLett.116.241102} {\bibfield  {journal} {\bibinfo  {journal}
  {Phys.\ Rev.\ Lett.}\ }\textbf {\bibinfo {volume} {116}},\ \bibinfo {pages}
  {241102} (\bibinfo {year} {2016}{\natexlab{g}})},\ \Eprint
  {http://arxiv.org/abs/1602.03840} {arXiv:1602.03840 [gr-qc]} \BibitemShut
  {NoStop}%
\bibitem [{\citenamefont {Varma}\ \emph {et~al.}(2019)\citenamefont {Varma},
  \citenamefont {Gerosa}, \citenamefont {Stein}, \citenamefont {Hebert},\ and\
  \citenamefont {Zhang}}]{Varma:2018aht}%
  \BibitemOpen
  \bibfield  {author} {\bibinfo {author} {\bibfnamefont {V.}~\bibnamefont
  {Varma}}, \bibinfo {author} {\bibfnamefont {D.}~\bibnamefont {Gerosa}},
  \bibinfo {author} {\bibfnamefont {L.~C.}\ \bibnamefont {Stein}}, \bibinfo
  {author} {\bibfnamefont {F.}~\bibnamefont {Hebert}}, \ and\ \bibinfo {author}
  {\bibfnamefont {H.}~\bibnamefont {Zhang}},\ }\href {\doibase
  10.1103/PhysRevLett.122.011101} {\bibfield  {journal} {\bibinfo  {journal}
  {Phys. Rev. Lett.}\ }\textbf {\bibinfo {volume} {122}},\ \bibinfo {pages}
  {011101} (\bibinfo {year} {2019})},\ \Eprint
  {http://arxiv.org/abs/1809.09125} {arXiv:1809.09125 [gr-qc]} \BibitemShut
  {NoStop}%
\bibitem [{\citenamefont {Blackman}\ \emph {et~al.}(2017)\citenamefont
  {Blackman}, \citenamefont {Field}, \citenamefont {Scheel}, \citenamefont
  {Galley}, \citenamefont {Ott}, \citenamefont {Boyle}, \citenamefont {Kidder},
  \citenamefont {Pfeiffer},\ and\ \citenamefont
  {Szilágyi}}]{Blackman:2017pcm}%
  \BibitemOpen
  \bibfield  {author} {\bibinfo {author} {\bibfnamefont {J.}~\bibnamefont
  {Blackman}}, \bibinfo {author} {\bibfnamefont {S.~E.}\ \bibnamefont {Field}},
  \bibinfo {author} {\bibfnamefont {M.~A.}\ \bibnamefont {Scheel}}, \bibinfo
  {author} {\bibfnamefont {C.~R.}\ \bibnamefont {Galley}}, \bibinfo {author}
  {\bibfnamefont {C.~D.}\ \bibnamefont {Ott}}, \bibinfo {author} {\bibfnamefont
  {M.}~\bibnamefont {Boyle}}, \bibinfo {author} {\bibfnamefont {L.~E.}\
  \bibnamefont {Kidder}}, \bibinfo {author} {\bibfnamefont {H.~P.}\
  \bibnamefont {Pfeiffer}}, \ and\ \bibinfo {author} {\bibfnamefont
  {B.}~\bibnamefont {Szilágyi}},\ }\href {\doibase 10.1103/PhysRevD.96.024058}
  {\bibfield  {journal} {\bibinfo  {journal} {Phys. Rev.}\ }\textbf {\bibinfo
  {volume} {D96}},\ \bibinfo {pages} {024058} (\bibinfo {year} {2017})},\
  \Eprint {http://arxiv.org/abs/1705.07089} {arXiv:1705.07089 [gr-qc]}
  \BibitemShut {NoStop}%
\bibitem [{\citenamefont {{LIGO Scientific Collaboration and Virgo
  Collaboration}}(2018)}]{GWOSC:GWTC}%
  \BibitemOpen
  \bibfield  {author} {\bibinfo {author} {\bibnamefont {{LIGO Scientific
  Collaboration and Virgo Collaboration}}},\ }\href@noop {} {\enquote {\bibinfo
  {title} {{GWTC-1}},}\ }\bibinfo {howpublished}
  {\href{https://doi.org/10.7935/82H3-HH23}{https://doi.org/10.7935/82H3-HH23}}
  (\bibinfo {year} {2018})\BibitemShut {NoStop}%
\bibitem [{\citenamefont {Verdinelli}\ and\ \citenamefont
  {Wasserman}(1995)}]{Verdinelli1995}%
  \BibitemOpen
  \bibfield  {author} {\bibinfo {author} {\bibfnamefont {I.}~\bibnamefont
  {Verdinelli}}\ and\ \bibinfo {author} {\bibfnamefont {L.}~\bibnamefont
  {Wasserman}},\ }\href {http://www.jstor.org/stable/2291073} {\bibfield
  {journal} {\bibinfo  {journal} {Journal of the American Statistical
  Association}\ }\textbf {\bibinfo {volume} {90}},\ \bibinfo {pages} {614}
  (\bibinfo {year} {1995})}\BibitemShut {NoStop}%
\bibitem [{\citenamefont {Ghosh}\ \emph {et~al.}(2016)\citenamefont {Ghosh}
  \emph {et~al.}}]{Ghosh:2016qgn}%
  \BibitemOpen
  \bibfield  {author} {\bibinfo {author} {\bibfnamefont {A.}~\bibnamefont
  {Ghosh}} \emph {et~al.},\ }\href {\doibase 10.1103/PhysRevD.94.021101}
  {\bibfield  {journal} {\bibinfo  {journal} {Phys. Rev. D}\ }\textbf {\bibinfo
  {volume} {94}},\ \bibinfo {pages} {021101(R)} (\bibinfo {year} {2016})},\
  \Eprint {http://arxiv.org/abs/1602.02453} {arXiv:1602.02453 [gr-qc]}
  \BibitemShut {NoStop}%
\bibitem [{\citenamefont {Ghosh}\ \emph {et~al.}(2018)\citenamefont {Ghosh},
  \citenamefont {Johnson-McDaniel}, \citenamefont {Ghosh}, \citenamefont
  {Mishra}, \citenamefont {Ajith}, \citenamefont {Del~Pozzo}, \citenamefont
  {Berry}, \citenamefont {Nielsen},\ and\ \citenamefont
  {London}}]{Ghosh:2017gfp}%
  \BibitemOpen
  \bibfield  {author} {\bibinfo {author} {\bibfnamefont {A.}~\bibnamefont
  {Ghosh}}, \bibinfo {author} {\bibfnamefont {N.~K.}\ \bibnamefont
  {Johnson-McDaniel}}, \bibinfo {author} {\bibfnamefont {A.}~\bibnamefont
  {Ghosh}}, \bibinfo {author} {\bibfnamefont {C.~K.}\ \bibnamefont {Mishra}},
  \bibinfo {author} {\bibfnamefont {P.}~\bibnamefont {Ajith}}, \bibinfo
  {author} {\bibfnamefont {W.}~\bibnamefont {Del~Pozzo}}, \bibinfo {author}
  {\bibfnamefont {C.~P.~L.}\ \bibnamefont {Berry}}, \bibinfo {author}
  {\bibfnamefont {A.~B.}\ \bibnamefont {Nielsen}}, \ and\ \bibinfo {author}
  {\bibfnamefont {L.}~\bibnamefont {London}},\ }\href {\doibase
  10.1088/1361-6382/aa972e} {\bibfield  {journal} {\bibinfo  {journal}
  {Classical Quantum Gravity}\ }\textbf {\bibinfo {volume} {35}},\ \bibinfo
  {pages} {014002} (\bibinfo {year} {2018})},\ \Eprint
  {http://arxiv.org/abs/1704.06784} {arXiv:1704.06784 [gr-qc]} \BibitemShut
  {NoStop}%
\bibitem [{\citenamefont {Vallisneri}\ \emph {et~al.}(2015)\citenamefont
  {Vallisneri}, \citenamefont {Kanner}, \citenamefont {Williams}, \citenamefont
  {Weinstein},\ and\ \citenamefont {Stephens}}]{Vallisneri:2014vxa}%
  \BibitemOpen
  \bibfield  {author} {\bibinfo {author} {\bibfnamefont {M.}~\bibnamefont
  {Vallisneri}}, \bibinfo {author} {\bibfnamefont {J.}~\bibnamefont {Kanner}},
  \bibinfo {author} {\bibfnamefont {R.}~\bibnamefont {Williams}}, \bibinfo
  {author} {\bibfnamefont {A.}~\bibnamefont {Weinstein}}, \ and\ \bibinfo
  {author} {\bibfnamefont {B.}~\bibnamefont {Stephens}},\ }\bibfield
  {booktitle} {\emph {\bibinfo {booktitle} {{Proceedings, 10th International
  LISA Symposium: Gainesville, Florida, USA, May 18-23, 2014}}},\ }\href
  {\doibase 10.1088/1742-6596/610/1/012021} {\bibfield  {journal} {\bibinfo
  {journal} {J. Phys. Conf. Ser.}\ }\textbf {\bibinfo {volume} {610}},\
  \bibinfo {pages} {012021} (\bibinfo {year} {2015})},\ \Eprint
  {http://arxiv.org/abs/1410.4839} {arXiv:1410.4839 [gr-qc]} \BibitemShut
  {NoStop}%
\bibitem [{\citenamefont {Abbott}\ \emph
  {et~al.}(2016{\natexlab{h}})\citenamefont {Abbott} \emph
  {et~al.}}]{TheLIGOScientific:2016wfe}%
  \BibitemOpen
  \bibfield  {author} {\bibinfo {author} {\bibfnamefont {B.~P.}\ \bibnamefont
  {Abbott}} \emph {et~al.} (\bibinfo {collaboration} {LIGO Scientific
  Collaboration, Virgo Collaboration}),\ }\href {\doibase
  10.1103/PhysRevLett.116.241102} {\bibfield  {journal} {\bibinfo  {journal}
  {Phys.~Rev.~Lett.}\ }\textbf {\bibinfo {volume} {116}},\ \bibinfo {pages}
  {241102} (\bibinfo {year} {2016}{\natexlab{h}})},\ \Eprint
  {http://arxiv.org/abs/1602.03840} {arXiv:1602.03840 [gr-qc]} \BibitemShut
  {NoStop}%
\bibitem [{\citenamefont {Abbott}\ \emph
  {et~al.}(2016{\natexlab{i}})\citenamefont {Abbott} \emph
  {et~al.}}]{TheLIGOScientific:2016zmo}%
  \BibitemOpen
  \bibfield  {author} {\bibinfo {author} {\bibfnamefont {B.~P.}\ \bibnamefont
  {Abbott}} \emph {et~al.} (\bibinfo {collaboration} {LIGO Scientific,
  Virgo}),\ }\href {\doibase 10.1088/0264-9381/33/13/134001} {\bibfield
  {journal} {\bibinfo  {journal} {Class. Quant. Grav.}\ }\textbf {\bibinfo
  {volume} {33}},\ \bibinfo {pages} {134001} (\bibinfo {year}
  {2016}{\natexlab{i}})},\ \Eprint {http://arxiv.org/abs/1602.03844}
  {arXiv:1602.03844 [gr-qc]} \BibitemShut {NoStop}%
\bibitem [{\citenamefont {Abbott}\ \emph
  {et~al.}(2016{\natexlab{j}})\citenamefont {Abbott} \emph
  {et~al.}}]{TheLIGOScientific:2016uux}%
  \BibitemOpen
  \bibfield  {author} {\bibinfo {author} {\bibfnamefont {B.~P.}\ \bibnamefont
  {Abbott}} \emph {et~al.} (\bibinfo {collaboration} {LIGO Scientific,
  Virgo}),\ }\href {\doibase 10.1103/PhysRevD.94.069903,
  10.1103/PhysRevD.93.122004} {\bibfield  {journal} {\bibinfo  {journal} {Phys.
  Rev.}\ }\textbf {\bibinfo {volume} {D93}},\ \bibinfo {pages} {122004}
  (\bibinfo {year} {2016}{\natexlab{j}})},\ \bibinfo {note} {[Addendum: Phys.
  Rev.D94,no.6,069903(2016)]},\ \Eprint {http://arxiv.org/abs/1602.03843}
  {arXiv:1602.03843 [gr-qc]} \BibitemShut {NoStop}%
\bibitem [{\citenamefont {Abbott}\ \emph
  {et~al.}(2016{\natexlab{k}})\citenamefont {Abbott} \emph
  {et~al.}}]{TheLIGOScientific:2016pea}%
  \BibitemOpen
  \bibfield  {author} {\bibinfo {author} {\bibfnamefont {B.~P.}\ \bibnamefont
  {Abbott}} \emph {et~al.} (\bibinfo {collaboration} {LIGO Scientific,
  Virgo}),\ }\href {\doibase 10.1103/PhysRevX.6.041015,
  10.1103/PhysRevX.8.039903} {\bibfield  {journal} {\bibinfo  {journal} {Phys.
  Rev.}\ }\textbf {\bibinfo {volume} {X6}},\ \bibinfo {pages} {041015}
  (\bibinfo {year} {2016}{\natexlab{k}})},\ \bibinfo {note} {[erratum: Phys.
  Rev.X8,no.3,039903(2018)]},\ \Eprint {http://arxiv.org/abs/1606.04856}
  {arXiv:1606.04856 [gr-qc]} \BibitemShut {NoStop}%
\bibitem [{\citenamefont {Veitch}\ \emph {et~al.}(2015)\citenamefont {Veitch},
  \citenamefont {Raymond}, \citenamefont {Farr}, \citenamefont {Farr},
  \citenamefont {Graff}, \citenamefont {Vitale}, \citenamefont {Aylott},
  \citenamefont {Blackburn}, \citenamefont {Christensen}, \citenamefont
  {Coughlin}, \citenamefont {Del~Pozzo}, \citenamefont {Feroz}, \citenamefont
  {Gair}, \citenamefont {Haster}, \citenamefont {Kalogera}, \citenamefont
  {Littenberg}, \citenamefont {Mandel}, \citenamefont {O'Shaughnessy},
  \citenamefont {Pitkin}, \citenamefont {Rodriguez}, \citenamefont {R\"over},
  \citenamefont {Sidery}, \citenamefont {Smith}, \citenamefont {Van Der~Sluys},
  \citenamefont {Vecchio}, \citenamefont {Vousden},\ and\ \citenamefont
  {Wade}}]{Veitch:2015}%
  \BibitemOpen
  \bibfield  {author} {\bibinfo {author} {\bibfnamefont {J.}~\bibnamefont
  {Veitch}}, \bibinfo {author} {\bibfnamefont {V.}~\bibnamefont {Raymond}},
  \bibinfo {author} {\bibfnamefont {B.}~\bibnamefont {Farr}}, \bibinfo {author}
  {\bibfnamefont {W.}~\bibnamefont {Farr}}, \bibinfo {author} {\bibfnamefont
  {P.}~\bibnamefont {Graff}}, \bibinfo {author} {\bibfnamefont
  {S.}~\bibnamefont {Vitale}}, \bibinfo {author} {\bibfnamefont
  {B.}~\bibnamefont {Aylott}}, \bibinfo {author} {\bibfnamefont
  {K.}~\bibnamefont {Blackburn}}, \bibinfo {author} {\bibfnamefont
  {N.}~\bibnamefont {Christensen}}, \bibinfo {author} {\bibfnamefont
  {M.}~\bibnamefont {Coughlin}}, \bibinfo {author} {\bibfnamefont
  {W.}~\bibnamefont {Del~Pozzo}}, \bibinfo {author} {\bibfnamefont
  {F.}~\bibnamefont {Feroz}}, \bibinfo {author} {\bibfnamefont
  {J.}~\bibnamefont {Gair}}, \bibinfo {author} {\bibfnamefont {C.-J.}\
  \bibnamefont {Haster}}, \bibinfo {author} {\bibfnamefont {V.}~\bibnamefont
  {Kalogera}}, \bibinfo {author} {\bibfnamefont {T.}~\bibnamefont
  {Littenberg}}, \bibinfo {author} {\bibfnamefont {I.}~\bibnamefont {Mandel}},
  \bibinfo {author} {\bibfnamefont {R.}~\bibnamefont {O'Shaughnessy}}, \bibinfo
  {author} {\bibfnamefont {M.}~\bibnamefont {Pitkin}}, \bibinfo {author}
  {\bibfnamefont {C.}~\bibnamefont {Rodriguez}}, \bibinfo {author}
  {\bibfnamefont {C.}~\bibnamefont {R\"over}}, \bibinfo {author} {\bibfnamefont
  {T.}~\bibnamefont {Sidery}}, \bibinfo {author} {\bibfnamefont
  {R.}~\bibnamefont {Smith}}, \bibinfo {author} {\bibfnamefont
  {M.}~\bibnamefont {Van Der~Sluys}}, \bibinfo {author} {\bibfnamefont
  {A.}~\bibnamefont {Vecchio}}, \bibinfo {author} {\bibfnamefont
  {W.}~\bibnamefont {Vousden}}, \ and\ \bibinfo {author} {\bibfnamefont
  {L.}~\bibnamefont {Wade}},\ }\href {\doibase 10.1103/PhysRevD.91.042003}
  {\bibfield  {journal} {\bibinfo  {journal} {Phys.\ Rev.\ D}\ }\textbf
  {\bibinfo {volume} {91}},\ \bibinfo {pages} {042003} (\bibinfo {year}
  {2015})},\ \Eprint {http://arxiv.org/abs/1409.7215} {arXiv:1409.7215 [gr-qc]}
  \BibitemShut {NoStop}%
\bibitem [{\citenamefont {Levinson}(1946)}]{Levinson:1947}%
  \BibitemOpen
  \bibfield  {author} {\bibinfo {author} {\bibfnamefont {N.}~\bibnamefont
  {Levinson}},\ }\href {\doibase 10.1002/sapm1946251261} {\bibfield  {journal}
  {\bibinfo  {journal} {Journal of Mathematics and Physics}\ }\textbf {\bibinfo
  {volume} {25}},\ \bibinfo {pages} {261} (\bibinfo {year} {1946})}\BibitemShut
  {NoStop}%
\bibitem [{\citenamefont {Durbin}(1960)}]{Durbin:1960}%
  \BibitemOpen
  \bibfield  {author} {\bibinfo {author} {\bibfnamefont {J.}~\bibnamefont
  {Durbin}},\ }\href {http://www.jstor.org/stable/1401322} {\bibfield
  {journal} {\bibinfo  {journal} {Revue de l'Institut International de
  Statistique / Review of the International Statistical Institute}\ }\textbf
  {\bibinfo {volume} {28}},\ \bibinfo {pages} {233} (\bibinfo {year}
  {1960})}\BibitemShut {NoStop}%
\bibitem [{\citenamefont {Jones}\ \emph {et~al.}(01  )\citenamefont {Jones},
  \citenamefont {Oliphant}, \citenamefont {Peterson} \emph
  {et~al.}}]{Jones:2001scipy}%
  \BibitemOpen
  \bibfield  {author} {\bibinfo {author} {\bibfnamefont {E.}~\bibnamefont
  {Jones}}, \bibinfo {author} {\bibfnamefont {T.}~\bibnamefont {Oliphant}},
  \bibinfo {author} {\bibfnamefont {P.}~\bibnamefont {Peterson}},  \emph
  {et~al.},\ }\href {http://www.scipy.org/} {\enquote {\bibinfo {title}
  {{SciPy}: Open source scientific tools for {Python}},}\ } (\bibinfo {year}
  {2001--}),\ \bibinfo {note} {[Online; accessed 16 April 2019]}\BibitemShut
  {NoStop}%
\bibitem [{\citenamefont {Welch}(1967)}]{Welch:1967use}%
  \BibitemOpen
  \bibfield  {author} {\bibinfo {author} {\bibfnamefont {P.}~\bibnamefont
  {Welch}},\ }\href {\doibase 10.1109/TAU.1967.1161901} {\bibfield  {journal}
  {\bibinfo  {journal} {IEEE Transactions on audio and electroacoustics}\
  }\textbf {\bibinfo {volume} {15}},\ \bibinfo {pages} {70} (\bibinfo {year}
  {1967})}\BibitemShut {NoStop}%
\bibitem [{\citenamefont {Prix}(2016)}]{bayesian_qnm}%
  \BibitemOpen
  \bibfield  {author} {\bibinfo {author} {\bibfnamefont {R.}~\bibnamefont
  {Prix}},\ }\href {https://dcc.ligo.org/LIGO-T1500618/public} {\emph {\bibinfo
  {title} {{Bayesian QNM search on GW150914}}}},\ \bibinfo {type} {Tech. Rep.}\
  \bibinfo {number} {LIGO-T1500618}\ (\bibinfo  {institution} {LIGO Scientific
  Collaboration},\ \bibinfo {year} {2016})\BibitemShut {NoStop}%
\bibitem [{\citenamefont {{Farr}}\ and\ \citenamefont
  {{Farr}}(2015)}]{kombine}%
  \BibitemOpen
  \bibfield  {author} {\bibinfo {author} {\bibfnamefont {B.}~\bibnamefont
  {{Farr}}}\ and\ \bibinfo {author} {\bibfnamefont {W.~M.}\ \bibnamefont
  {{Farr}}},\ }\href {https://github.com/bfarr/kombine} {\enquote {\bibinfo
  {title} {kombine: a kernel-density-based, embarrassingly parallel ensemble
  sampler},}\ } (\bibinfo {year} {2015}),\ \bibinfo {note} {in
  prep}\BibitemShut {NoStop}%
\end{thebibliography}%
    \end{document}
\fi

}{
}

\end{document}